\documentclass[a4paper,12pt]{article}
\pdfoutput=1
\usepackage{amssymb}
\usepackage{amsmath}
\usepackage{epsfig}
\usepackage{graphicx}
\usepackage{slashed}
\usepackage{xcolor,color}
\usepackage{ulem}
\usepackage{subfigure}
\usepackage[toc,page]{appendix}
\usepackage[makeroom]{cancel}

\makeatletter

\@addtoreset{equation}{section}
\makeatother
\def\be{\begin{equation}}
\def\ee{\end{equation}}
\def\bea{\begin{eqnarray}}
\def\eea{\end{eqnarray}}
\def\bgn{\begin{align}}
\def\egn{\end{align}}

\def\({\left(}
\def\){\right)}
\def\<{\left<}
\def\>{\right>}

\def\({\left(}
\def\){\right)}
\def\<{\left<}
\def\>{\right>}
\def\!{\right|}
\def\|{\left|}

\def\[{\left[}
\def\]{\right]}

\def\+{\bar}

\def\r{\mathring}

\def\nt{{\negthinspace}}

\usepackage{hyperref}
\hypersetup{colorlinks,%
  citecolor=blue,%
  linkcolor=blue,%
  pdftex}

\begin{document}

\begin{titlepage}
\vskip1cm
\begin{flushright}
\end{flushright}
\vskip0.25cm
\centerline{
\bf \large 
Python's Lunches in Jackiw-Teitelboim gravity with matter
} 
\vskip0.8cm \centerline{ \textsc{
 Dongsu Bak,$^{ \negthinspace  a}$  Chanju Kim,$^{ \negthinspace b}$ Sang-Heon Yi,$^{\negthinspace c}$ Junggi Yoon$^{\, d,e,f}$} }
\vspace{0.8cm} 
\centerline{\sl  a) Physics Department \& Natural Science Research Institute}
\centerline{\sl University of Seoul, Seoul 02504 \rm KOREA}
 \vskip0.2cm
 \centerline{\sl b) Department of Physics, Ewha Womans University,
  Seoul 03760 \rm KOREA}
   \vskip0.2cm
 \centerline{\sl c) Center for Quantum Spacetime \&  Physics Department,
  Sogang University,  Seoul 04107 \rm KOREA}
   \vskip0.2cm   
 \centerline{\sl d) Asia Pacific Center for Theoretical Physics, Pohang 37673 \rm KOREA}
   \vskip0.2cm
 \centerline{\sl e) Department of Physics, POSTECH, Pohang 37673 \rm KOREA}
 \centerline{\sl f) School of Physics, Korea Institute for Advanced Study, Seoul 02455, Korea}
\vskip0.4cm

 \centerline{
\tt{(\small dsbak@uos.ac.kr,\,cjkim@ewha.ac.kr,\,shyi@sogang.ac.kr,\,junggi.yoon@apctp.org})} 
  \vspace{1.5cm}
\centerline{ABSTRACT} \vspace{0.65cm} 
{
\noindent 
We study Python's lunch geometries in the two-dimensional Jackiw-Teitelboim model
coupled to a massless scalar field in the semiclassical limit. We show that 
all extrema including the minimal quantum extremal surface, bulges and appetizers 
lie inside the horizon. We obtain fully back-reacted general bulk solutions
with a massless scalar field, which can be understood as deformations
of black holes. The temperatures of the left/right black holes become
in general different from each other. Moreover, in the presence of both 
state and source deformations at the same time, the asymptotic black hole 
spacetime is further excited from that of the vacuum solution. 
We provide information-theoretic interpretation of deformed geometries including 
Python's lunches, minimal quantum extremal surface and appetizers according to the entanglement wedge 
reconstruction hypothesis. By considering the restricted circuit complexity 
associated with Python's lunch geometries and the operator complexity 
of the Petz map reconstructing a code space operation, we show that 
the observational probability of Python's lunch degrees of freedom 
from the boundary is exponentially suppressed. Thus, any bulk 
causality violation effects related with Python's lunch degrees are 
suppressed nonperturbatively.
}

\end{titlepage}


\section{Introduction
}\label{sec1}

The concept of entropy plays a fundamental role in 
various   fields of theoretical physics.
Especially in black hole physics,
the entropy 
may pave a door-way to our understanding 
of quantum gravity. 
One of the well-established results in this regard 
 is the 
   Bekenstein-Hawking area 
formula~\cite{Bekenstein:1973ur,Hawking:1974rv}, which says 
that the black hole entropy 
is given by its horizon area divided by $4 G$. 
Though this formula is regarded 
as a clue to the holographic nature of underlying 
degrees of freedom, it does not reveal 
their precise microscopic origin  unfortunately. 
 Based on the fact that the horizon works 
 as a one-way door 
(not allowing anything to escape from a black hole classically),  it has been proposed that the entropy may be understood as 
a  fine-grained entanglement entropy between the interior and the exterior regions, 
divided by the black-hole horizon
~\cite{Bombelli:1986rw,Srednicki:1993im}.

As was argued by Page some time ago~\cite{Page:1993df}, the entanglement entropy of black holes should respect the unitary evolution of quantum mechanics and follow the so-called Page curve.  This  is incompatible with the result that the Hawking radiation is 
 Planckian~\cite{Hawking:1974sw} while any correction to this result should be exponentially suppressed~\cite{Mathur:2009hf} and so an initially pure state is evolved into a mixed, as far as there is no remnant in the black hole evaporation process and the low energy effective field theory approach is valid. This conflict is one way to explain the information loss paradox in black hole physics, which is sharply reformulated as the firewall paradox~\cite{Almheiri:2012rt}.  For old black holes which have emitted more than half of their mass  through the Hawking radiation, a newly generated Hawking radiation is 
entangled with both the early radiation and the behind horizon degrees of freedom, 
which is claimed to be in contradiction with the monogamy 
of the entanglement. 
Thus this conclusion  must be evaded in some ways. 

In this regard, there were  rather crucial 
developments based on the so-called generalized entropy~\cite{Bekenstein:1973ur,Bombelli:1986rw}. One begins with a boundary region ${\cal I}$ and its complement  on a given boundary time slice. 
Along with
 a  codimension two bulk surface $P$ that is homologous to ${\cal I}$ and divides a bulk Cauchy slice ending on the boundary time slice 
 into two regions, 
the generalized entropy $S_P$ can be defined, which consists of two contributions;
One is given by the area of $P$ (divided by $4G$) and the other by 
the entanglement entropy of bulk fields between the two regions divided by the surface $P$. 
With this preliminary, the procedure to obtain the minimal quantum extremal surface (QES$_{\rm min}$) goes as follows; First extremize $S_P$ spatially on a Cauchy slice specified in the above\footnote{Usually, one picks all local minima of $S_P$ at this first stage.  In this note, we instead include all extrema, which leads to the same QES$_{\rm min}$ at the end of the day.}, and then maximize 
 among all possible choices of Cauchy slices for the given boundary, 
which leads to 
a set of extremal surfaces.  Among these extremal surfaces, we choose the global minimum, which singles out the desired  QES$_{\rm min}$~\cite{Wall:2012uf,Engelhardt:2014gca,Akers:2019lzs}. 
This 
can be thought of as a generalization of the Ryu-Takayanagi surface~\cite{Ryu:2006ef,Ryu:2006bv,Hubeny:2007xt}, which has been explored in various contexts as a holographic realization of the entanglement entropy of a boundary region. 
The generalized entropy in some examples can be computed explicitly, explains the Page curve, and evades the firewall  through some further identification: entanglement wedge island. It has been argued  under the central dogma that some part of black hole 
interior degrees of freedom associated with the island are actually belonging to the radiation (see for a review~\cite{Almheiri:2020cfm}); 
An outside
observer acting on the radiation only may in principle probe the island degrees of freedom directly,
which appears rather bizarre 
as they are  at least shielded by the horizon!

Python's lunches are closely related to the island wedges just mentioned~\cite{Brown:2019rox}. In general, among all extremal surfaces, one may have a local maximum placed between two saddle
surfaces. Such a local maximal surface is called a bulge surface whereas the remaining saddle surfaces except the QES$_{\rm min}$ are called appetizers. Python's lunch geometries  are involving   at least 
one bulge and one appetizer. 
In this note we would like to understand the nature of Python's lunch degrees of freedom. 
In order to understand them, one may borrow various concepts and results from quantum 
information theory and its realization in the context of the AdS/CFT correspondence. These
include the entanglement wedge reconstruction, quantum error correcting codes, the holographic tensor network, 
 circuit complexity
and so on~\cite{Swingle:2009bg,Czech:2012bh,Headrick:2014cta,Almheiri:2014lwa,Jafferis:2015del,Dong:2016eik,Faulkner:2017vdd,Cotler:2017erl,Chen:2019gbt}.

In particular, quantum complexity plays an essential role in our understanding of the Python's lunch degrees of freedom. It has been known that there are two versions of complexity. 
One is the unrestricted version of complexity associated with an observer who may access the entire quantum system of interest. For instance, in two-sided eternal black holes, probed by
an observer who may access both boundaries, 
the unrestricted complexity of the thermo-field-double (TFD) state under time evolution is holographically related to the time-dependent expansion of the black hole interior region~\cite{Susskind:2014rva,Brown:2015lvg,Brown:2015bva}.  In the bulk picture, this complexity may be identified using the 
volume/action conjecture. The other is the restricted complexity  where an associated 
observer can access only part of the underlying quantum system. A rather well-known example
is the computational complexity arising in decoding  information out of Hawking radiation \cite{Hayden:2007cs,Harlow:2013tf}. In this note, we shall be mainly concerned about the restricted complexity in order to understand the nature of Python's lunch degrees of freedom.

In this note we focus on the two-dimensional Jackiw-Teitelboim (JT) model~\cite{Jackiw:1984je,Teitelboim:1983ux} coupled to massless scalar fields in its semiclassical limit. 
We shall identify all possible deformations of black hole solutions which may involve Python's lunch geometries. It turns out that, in the semiclassical limit,  all extrema including QES$_{\rm min}$, bulges and appetizers lie inside the horizon, which basically follows from  the null energy condition (NEC). 
This deformed two-sided black hole spacetime is to be divided into left and right wedges 
that share QES$_{\rm min}$ as a separating surface (point in 2d). Under the entanglement wedge reconstruction hypothesis, all  local bulk operations within the left/right wedge can be reconstructed out of the corresponding operations on the left/right boundary 
 Hilbert space. For definiteness, let us choose the 
right wedge 
and  consider a minimal Python's  lunch lying behind the horizon of the right side. Within this,  there are specific local bulk degrees of freedom that are genuinely associated with the Python's lunch ($\alpha$-degrees in short) and the remaining local bulk degrees of freedom in the region outside the appetizer ($i$-degrees in short). The combined bulk Hilbert space should be mapped to the full boundary Hilbert space of the right side.
 We shall review the restricted 
complexity 
of Python's lunch geometries that is exponential in their  number of degrees. We shall also review the operational complexity of the Petz map by which one reconstructs
code space operations acting on the bulk Python's lunch Hilbert space  out of $i$-Hilbert space of a one-sided boundary.  Based on these, we shall argue that observational probability of Python's lunch 
degrees of freedom by some boundary experiment acting on $i$-Hilbert space is  exponentially suppressed. Thus, any bulk causality violation effects related with Python's lunch degrees are suppressed nonperturbatively.

This paper is organized as follows. In section \ref{sec2}  our setup is given  with the emphasis on the two-sided black hole solutions. 
In section \ref{sec3} we  provide a generic deformation of a scalar field solving the equation of motion. For concreteness we consider the massless scalar field only. We also present the 
general
profiles of the dilaton field.
In section \ref{sec4} it is shown that the deformation of the dilaton field solution can be understood as the deformation of black holes. Exploring the effects of the scalar field deformation on the black holes, we show that the   temperatures of two-sided black holes can be asymmetric between the left and right sides.  The end points of the horizon are also studied and it is shown that the traversable wormholes cannot be formed by our deformation. 
We show that QES$_{\rm min}$, Python's lunches, and appetizers reside behind the horizon in our semiclassical setup.
In section \ref{sec5} 
we provide some geometric interpretation of our results including python's lunch, appetizer, QES$_{\rm min}$,  
 and the entanglement wedge reconstruction hypothesis. In section \ref{sec6},  based on our study,  we give some reviews and comments on the related aspects of the complexity, the information  reconstruction  from the one-sided boundary state,
and the interaction among the 
degrees of freedom including 
Python's lunch degrees of freedom. In the final section, we summarize our results and provide some future directions.

\section{Two-dimensional dilaton gravity 
}\label{sec2}

Our basic setup is a 2d dilaton gravity  known as the JT  
model coupled to a matter field~\cite{Jackiw:1984je,Teitelboim:1983ux,Almheiri:2014cka}, whose action is given by
\bea
I=I_{top}+{1\over 16\pi G}\int_M d^2 x \sqrt{-g}\, \phi \left( R+\frac{2}{\ell^2}\right) +I_M(g, \chi)\,,
\label{euclidaction}
\eea
where 
\begin{align}    \label{}
 I_{top}&=  {\phi_0\over 16\pi G}\int_M d^2 x \sqrt{-g}  R\,,  \nonumber \\
I_M &= -\frac{1}{2}\int_M d^2 x \sqrt{-g} \left( g^{ab}\nabla_a \chi 
\nabla_b \chi + m^2 \chi^2 \right)\,.
\end{align}
For a well-defined variational principle, 
we need to add a surface term 
\begin{equation} \label{}
I_{surf} =   {1\over 8\pi G}\int_{\partial M} \sqrt{-\gamma}\, (\phi_0 + \phi ) \, K 
\end{equation}
to the above 
where $\gamma_{ij}$ and $K$  
denote the induced metric and  the extrinsic curvature on the boundary $\partial M$ which is taken to be timelike.

The equations of motion read
\begin{align}   
&R+\frac{2}{\ell^2}=0\,, \label{metric} \\
& \nabla^2 \chi -m^2 \chi =0\,,   \label{phieq} \\
& \nabla_a \nabla_b \phi -g_{ab} \nabla^2 \phi + \frac{1}{\ell^{2}}g_{ab} \phi = - 8 \pi G T_{ab}\,,
\end{align}
where
\begin{equation} \label{}
T_{ab} = \nabla_a \chi \nabla_b \chi  -\frac{1}{2} g_{ab} \left( g^{cd}\nabla_c \chi 
\nabla_d \chi + m^2 \chi^2 \right)\,.
\end{equation}
The  global AdS$_{2}$ space corresponds to  the metric solution to the equations of motion,  whose form is given by
\begin{equation} \label{Gmetric}
ds^2 =\frac{\ell^2}{\cos^2 \mu} \left(-d\tau^2 + d\mu^2  \right)\,,
\end{equation}
where the spatial coordinate $\mu$ is ranged over $[-\frac{\pi}{2},\frac{\pi}{2}]$. The dilaton field configuration for the most general vacuum black hole solution   reads 
\begin{equation}
\phi= \phi_{BH}(L,b,\tau_{B})\equiv \bar\phi \, L\,\,\frac{(b+b^{-1}) \cos (\tau-\tau_B) -(b-b^{-1}) \sin \mu}{2 \cos \mu} \,.
\label{dilaton}
\end{equation}
%
By the coordinate transformation 
\begin{align}    \label{}
\frac{r}{L} &= \frac{(b+b^{-1}) \cos (\tau-\tau_B) -(b-b^{-1}) \sin \mu}{2\cos \mu}\,,  \nonumber \\
 \tanh \frac{t L }{\ell^2} &=\frac{2\sin (\tau-\tau_B)}{(b+b^{-1}) \sin \mu -(b-b^{-1}) \cos (\tau-\tau_B)}\,,
 \label{coorb}
\end{align}
one is led to the corresponding AdS black hole metric 
\begin{equation} \label{btz}
ds^2= - \frac{r^2-L^2}{\ell^2} dt^2+ \frac{\ell^2}{r^2-L^2} dr^2\,,
\end{equation}
with
\begin{equation} \label{}
\phi= \bar\phi \, r\,.
\end{equation}
%
In general, this  black hole metric describes only the Rindler wedge of  two-sided AdS black holes. The location of singularity is defined by the curve $\Phi^2\equiv \phi_0 +\phi =0$ in 
the above dilaton field, and  $\Phi^2$ might be viewed as characterizing the size of  transverse space  from the viewpoint of dimensional reduction from higher dimensions\footnote{For example,  the dimensional reduction of 4d near-extremal black holes has been investigated in  \cite{Nayak:2018qej}, which leads to our  JT gravity with matter fields.  This reduction is well controlled  once $\phi|_{\rm cutoff} \ll \phi_0$ \cite{Maldacena:2016upp,Nayak:2018qej} where $\phi|_{\rm cutoff}$ denotes the value of the dilaton at asymptotic cut-off trajectories introduced below.
Of course, for the well-defined nearly AdS$_2$ dynamics, we also require $\bar\phi\, \ell \ll \phi|_{\rm cutoff}$. With these conditions, our solutions below can be consistently embedded into the 4d theory.  
} \cite{Almheiri:2014cka}.  
%
%
\def\xxx1{
\begin{figure}
\vskip-1cm
\begin{center}
\includegraphics[width=6.3cm,clip]{etbn.pdf}
\end{center}
\vskip-1cm
\caption{
\label{fig02} We draw the Penrose diagram for the AdS$_2$ black hole with $b=1$ in $(\tau,\mu)$ space where the wiggly red lines represent the location 
of singularity.
}
\end{figure}
}
In this left/right symmetric two-sided black hole,  one can see that
the Hawking temperature, the entropy  and energy  are given by 
\begin{equation} \label{}
T= \frac{1}{2\pi} \frac{L}{\ell^2}\,,  \qquad S= S_0 +{\cal C} T\,,   \qquad 
E = \frac{1}{2} {\cal C}T^2\,,
\end{equation}
where  $S_{0}= \frac{\phi_0} {4G}$ and ${\cal C} =\frac{\pi \bar\phi \ell^2}{2 G}$ denote the ground state entropy and the specific heat, respectively. In general, these physical quantities may be different for left/right Rindler wedges  in deformed two-sided black holes. 
In the next sections  we  consider general deformations with a scalar field  
starting from the left-right symmetric  black hole configuration with
$\tau_B=0$ and $b=1$. 
We find that the above left-right thermodynamic
 quantities indeed become different from 
each other in general with a nontrivial  matter field profile. 

\begin{figure}[htbp]   
\begin{center}
\includegraphics[width=0.35\textwidth]{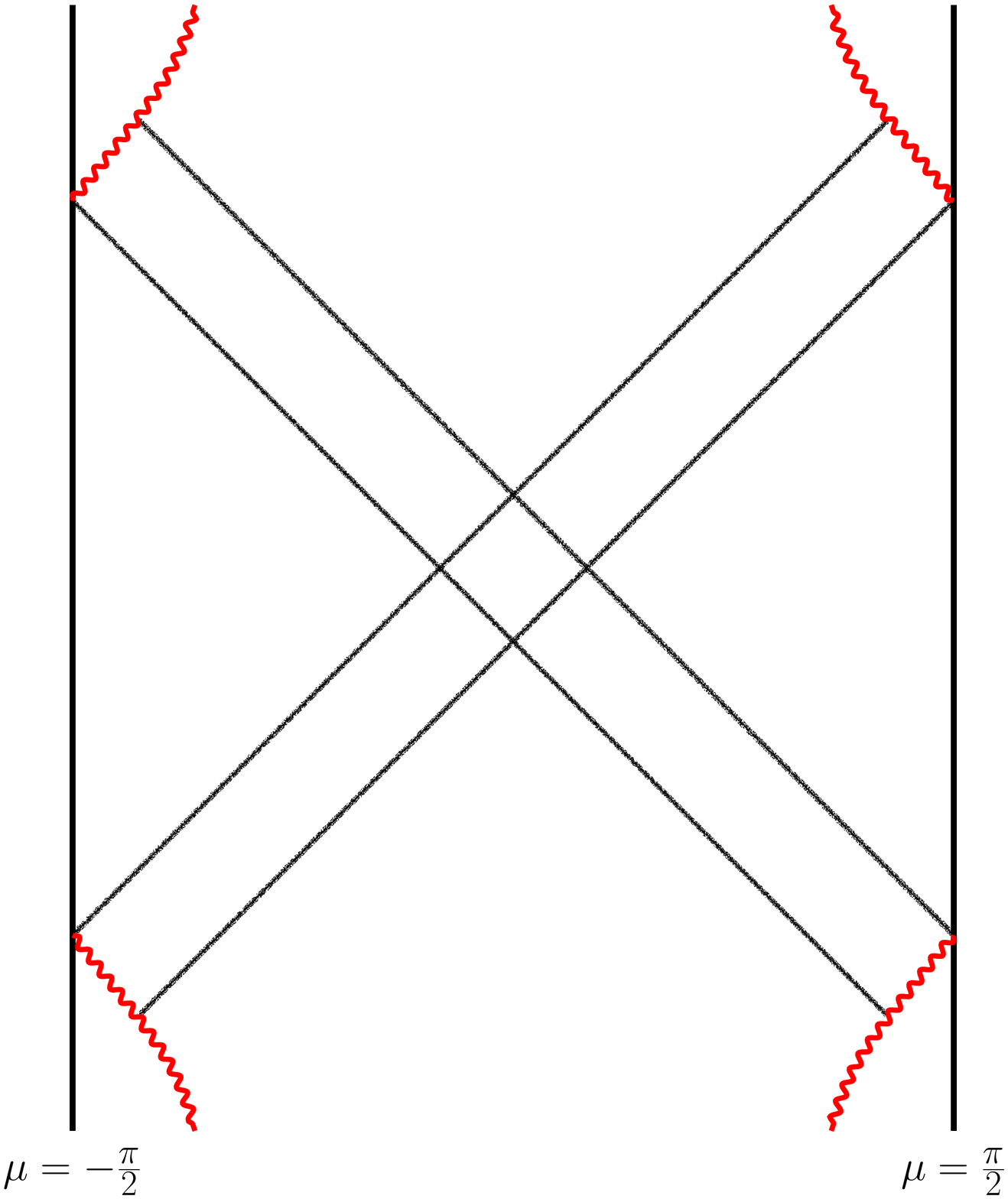}
\caption{The Penrose diagram for the Janus deformation is depicted.}
\label{Janus}
\end{center}
\end{figure}

Before going ahead, let us recall a simple left-right asymmetrically deformed configuration with
a massless scalar, known as the eternal Janus deformation~\cite{Bak:2007jm,Bak:2018txn}. In this case, the black hole temperature and  the dilaton field configuration are left-right symmetric under the exchange $\mu\leftrightarrow -\mu$, but the matter field configuration is not left-right symmetric. This simple Janus solution is given by
\begin{equation} \label{janusdef}
\chi =\gamma (\mu-\kappa_{0})\,, \qquad \phi = \bar{\phi}L_{0}\frac{\cos\tau}{\cos\mu} -4\pi G\gamma^{2}(1+\mu\tan\mu)\,,
\end{equation}
%
The Penrose diagram of this configuration with $\phi_{0}$ contribution  is given in Figure \ref{Janus}.
%
%

Now, we would like to present the matter field   solution to  the equations of motion in  (\ref{phieq}), which is given by
\begin{equation} \label{}
\chi= \sum^\infty_{n=0} c_n \, {\cal N}_n \cos^{\cal D} \mu \, C_n^{\cal D} (\sin \mu) \, e^{-i(n+{\cal D})\tau 
}  + {\rm c.c.}\ ,
\end{equation}
where 
\begin{equation} \label{}
{\cal N}_n =2^{{\cal D} -1} \Gamma({\cal D}) {\textstyle  \sqrt{\frac{\Gamma(n+1)}{\pi \Gamma(n+2{\cal D})}}}\,,
\end{equation}
and $C^{\cal D}_n(x)$ denotes the Gegenbauer polynomial. 
 Here, the parameter ${\cal D}$ is related to  the mass of the scalar field $\chi$ as
\begin{equation} \label{}
{\cal D}=\Delta_\pm= \frac{1}{2} \left( 1 \pm \sqrt{1+4m^2}\right)\,.
\end{equation}
The AdS/CFT dictionary tells us that  the  bulk matter field  $\chi$ with mass $m$, when $m^2 \ge 0$,   is dual  
to a scalar primary operator $O_\Delta (t)$, where its dimension $\Delta$ is  given by the larger one of $\Delta_{\pm}$ as 
\begin{equation} \label{}
\Delta = \Delta_{+}=\frac{1}{2} \left( 1 + \sqrt{1+4m^2}\right)\,,
\end{equation}
The deformation by the  scalar field $\chi$  with ${\cal D}=\Delta_{+}$  corresponds to a state deformation of the dual field theory, while the other  deformation  with ${\cal D}=\Delta_{-}$ does to a source deformation.   When scalar field mass squared  is negative but above the Breitenlohner-Freedman bound~\cite{Breitenlohner:1982jf}, which is given by $ 0 > m^2  > -1/4 $ in our case,   both possibilities of operator dimensions may be realized as 
\bea
\Delta =\Delta_\pm= \frac{1}{2} \left( 1 \pm \sqrt{1+4m^2}\right)\,.
\label{deltapm}
\eea

In this paper, we focus on the bulk classical configuration by ignoring the loop corrections. This limited consideration may be justified by keeping the Newton constant very small and by introducing  a controllable  number of identical scalar fields, say $K$,  without self-interactions~\cite{Maldacena:2018lmt}. With the combination $K G$ fixed, one finds that any loop corrections are suppressed
as $\mathcal O(1/K^n)$ where $n$ is counting the number of loops involved. 

\section{General deformations with a massless 
scalar field
}\label{sec3}

From now on, we shall focus on turning on a massless scalar field 
that is 
dual to a scalar operator of dimension $\Delta=1$.  In this section we are mainly  interested in fully 
back-reacted bulk solutions with the massless scalar field. Based on these solutions, 
we shall identify rather general Python's lunch geometries that are arising if certain parameter conditions are met. 
For this, we  note that the metric solution to the equation~\eqref{metric} remains the same while  the bulk scalar equation 
(\ref{phieq}) (with $m^2=0$) may be solved explicitly by
\bea\label{fulldefor} 
\chi = \chi_v +\chi_s\ ,
\eea
where
\be \label{chiv}
\chi_v=\sum^\infty_{n=1}a_{n}\sin n\big(\mu+{\frac{\pi}{2}}\big) \cos n(\tau-\tau^v_n) \ ,
\ee 
and 
\be \label{chiv}
\chi_s=b_0 +\sum^\infty_{n=1}b_{n}\cos n\big(\mu+{\frac{\pi}{2}}\big) \cos n(\tau-\tau^s_n) \ .
\ee 
This describes a full-set of classical solutions of the scalar equation in the ambient 
AdS$_2$ space with 
the global coordinates. 
The bulk solution with $\chi=\chi_v$ corresponds to the state deformation of the left/right boundary quantum system without 
deforming our boundary theories\footnote{One may specify the well-known correspondence between  JT gravity and SYK system. But such detailed specification does not play much role in our  description of this paper.}; This leads to a  nontrivial vev of the dimension-one operator $O$ in the boundary theory. 
On the other hand, the bulk solution with $\chi=\chi_s$ corresponds to the source deformation of boundary Lagrangian $L_0$ by $L_0 + s_{L/R}(u)O(u)$ where $u$ denotes the time coordinate of our left/right boundary system 
going upward direction\footnote{Note that the left-right values of vev  and source are not  independent since our undeformed bulk configuration corresponds to the TFD state in the boundary.}.     

In our JT model, the metric remains to be AdS$_2$ with an arbitrary matter perturbation, as mentioned previously. For the dilaton field, we shall consider deformations from the vacuum 
solution
\be 
\phi_{bg} =\, \bar\phi L_0 \, \frac{\cos \tau}{\cos \mu}\,,
\ee
which describes a two-sided black hole with temperature $T_0=\frac{1}{2\pi}\negthinspace\frac{L_0}{\ell^2}$. When the matter field is turned on, the two-sided black hole will be perturbed  in general  left-right asymmetrically 
as we shall analyze in detail in later sections.

Considering first the state perturbation with $\chi=\chi_v$, the dilaton solution becomes
\be \label{fullsoln}
\phi = \phi_{bg} 
+8\pi G  \sum^\infty_{m=1}\sum^\infty_{n=1}a_{m}a_{n} \phi^v_{m,n} \ ,
\ee
where
\begin{align}\label{diagn}
\phi^v_{n,n} &=\frac{(-1)^n\, n}{8(4n^2-1)}
\left( 2n{\cos 2n \mu}+\sin 2n \mu \tan \mu\right)\cos 2n (\tau-\tau^v_n)
-\frac{n^2}{4}(1+\mu\tan \mu)\ ,
\end{align}
and
\begin{align}\label{off1n}
&\phi^v_{n,n+1}=\phi^v_{n+1,n}\nonumber\\ \nonumber
 &=\frac{(-1)^{n+1}\sec\mu}{16(2n+1)}
\Big[ (n+1){\sin 2n \mu}+n\sin (2n+2) \mu \Big]\cos[(2n+1)\tau\nt-\nt (n+1)\tau^v_{n+1} \nt-\nt n  \tau^v_n]\\
& \ \ \ -\frac{n(n+1)}{{8}}(\sin \mu+\mu\sec \mu)\cos[\tau\nt-\nt (n+1)\tau^v_{n+1} \nt+\nt n  \tau^v_n]\ ,
\end{align}
with $n=1,2,\cdots$.
The remaining off-diagonal contribution 
 $\phi^v_{m,n}\,\, (m\neq n,n\pm1)$  is given by
\begin{align}
\phi^v_{m,n} =&\frac{mn}{8\cos \mu}
\left[
\cos[n(\tau\nt-\nt\tau^v_n)\nt-\nt m(\tau\nt -\nt \tau^v_m)]  \textstyle{\left( \frac{\sin(n-m+1)(\mu+\frac{\pi}{2})}{(n-m+1)(n-m)}-\frac{\sin(n-m-1)(\mu+\frac{\pi}{2})}{(n-m-1)(n-m)}\right)}
 \right. 
 \nonumber\\
&\left.
+\cos[n(\tau\nt-\nt\tau^v_n)\nt+\nt m(\tau\nt -\nt \tau^v_m)] \textstyle{\left( \frac{\sin(n+m+1)(\mu+\frac{\pi}{2})}{(n+m+1)(n+m)}-\frac{\sin(n+m-1)(\mu+\frac{\pi}{2})}{(n+m-1)(n+m)}\right)}
 \right]\,,
\end{align} 
which satisfies $\phi^v_{m,n}=\phi^v_{n,m}$.
The terms in (\ref{diagn}) and (\ref{off1n}) can be obtained by taking
   $\lim_{m\rightarrow n}\phi^v_{m,n}$ and $\lim_{m\rightarrow n+1}\phi^v_{m,n}$
up to some homogeneous terms. 

For the sake of an illustration, we here present the gravity solution with
$\chi=a_1 \chi^v_1+a_2 \chi^v_2+a_3 \chi^v_3$ where 
$\chi_n^v=\sin n(\mu+{\frac{\pi}{2}}) \cos n(\tau-\tau^v_n) $.
The corresponding dilaton solution in this case becomes
\be   \label{pv123}
\phi = \phi_{bg} +8\pi G \left(a_1^2\phi^v_{1,1}+ a_2^2 \phi^v_{2,2}+a_3^2\phi^v_{3,3} + 2a_1 a_2 \phi^v_{1,2}+ 2 a_1 a_3 \phi^v_{1,3}+ 2a_2 a_3 \phi^v_{2,3}\right)\,,
\ee
where the three diagonal terms are given by
\begin{align} \label{phiv123}
\phi^v_{1,1}\, &= -  
\left[ 
\frac{1}{12} \cos^2 \mu \cos 2 (\tau-\tau^v_1) 
   + \frac{1}{4} (1+ \mu \tan \mu) 
\right]\,,
\nonumber\\
\phi^v_{2,2}\,&= -  
\left[ 
\frac{1}{15} (2-3\cos 2\mu)\cos^2 \mu \cos 4 (\tau-\tau^v_2) 
   +  (1+ \mu \tan \mu) 
\right]\,,
\\
\phi^v_{3,3}\, &= -\frac{3}{140} \Bigl[ \cos^2 \mu(9 - 16 \cos2 \mu + 10 \cos4 \mu) \cos
     6 (\tau -  \tau^v_3) + 105 (1 + \mu \tan \mu)\Bigr] \ .\nonumber
\end{align}
The remaining off-diagonal terms can be organized as
\begin{align}\label{phic123}
\phi^v_{1,2} &=  \negthinspace 
\frac{1}{6} \cos^2 \mu \sin \mu \cos  (3\tau\negthinspace-\negthinspace\tau^v_1\negthinspace-\negthinspace2\tau^v_2) 
   \negthinspace-\negthinspace \frac{1}{4}\left(\frac{\mu}{ \cos \mu}\negthinspace +\negthinspace\sin \mu\right) \cos(\tau \negthinspace+\negthinspace\tau^v_1\negthinspace-\negthinspace 2\tau^v_2) \ ,\nonumber\\
\phi^v_{2,3}\, &=  
-\frac{1}{80} \Bigl[(60 \cos(\tau + 2 \tau^v_2 - 3 \tau^v_3)
\Bigl(\frac{\mu}{\cos\mu} +  \sin\mu\Bigr) \nonumber\\ &\ \ - 
 \cos(5\tau - 2 \tau^v_2 - 3 \tau^v_3) 8 \cos^2\mu (-1 + 4 \cos2 \mu) \sin\mu)
\Bigr]\,,
\nonumber\\
\phi^v_{1,3}\, &=-  
\frac{1}{20} \cos^2 \mu\Bigl[5 \cos(2 \tau + \tau^v_1 - 3 \tau^v_3) + (2 - 3 \cos 2 \mu) \cos(4 \tau - \tau^v_1 - 3 \tau^v_3)
\Bigr]\,.
\nonumber
\end{align}
Note that the left and right boundaries of  our AdS$_2$ spacetime are located respectively 
at $\mu=\pm\pi/2$. 
In these asymptotic regions, the dilaton solution 
behaves as
\begin{equation} \label{phias}
\phi \rightarrow \bar\phi L_0 \frac{Q(\tau)}{\cos \mu} + O(\cos\mu) \,.
\end{equation}
Indeed one may explicitly check this behavior for $\phi^v_{1,1},\, \phi^v_{2,2},\, \phi^v_{3,3}$ and
 $\phi^v_{1,2},\, \phi^v_{2,3}$ with nonvanishing contributions to $Q(\tau)$. The remaining term  $\phi^v_{1,3}\rightarrow  O(\cos^2\mu) $ as we approach the left/right boundaries does not contribute to $Q(\tau)$. This trend continues with
the general mode-solution $\phi^v_{m,n}$ for arbitrary $m$ and $n$. 
Namely, the $Q(\tau)$ contributions from  $\phi^v_{n,n}$ and $\phi^v_{n,n+1}(= \phi^v_{n,n+1})$ 
are nonvanishing. On the other hand,
 $\phi^v_{m,n} (m\neq n, n\pm 1)
\rightarrow  O(\cos^2\mu) $ as $\mu \rightarrow \pm\pi/2$ 
and, consequently, has no contribution to $Q(\tau)$.

We here also  present a general solution for the source deformation with $\chi=\chi_s$. 
The corresponding dilaton solution can be obtained as
\be \label{fullsol}
\phi = \phi_{bg} 
+8\pi G  \sum^\infty_{m=1}\sum^\infty_{n=1}b_{m}b_{n} \phi^s_{m,n} \ ,
\ee
where
\begin{align}\label{diag}
\phi^s_{n,n} &=\frac{(-1)^{n+1}\, n}{8(4n^2-1)}
\left( 2n{\cos 2n \mu}+\sin 2n \mu \tan \mu\right)\cos 2n (\tau-\tau^s_n)
-\frac{n^2}{4}(1+\mu\tan \mu)\ ,
\end{align}
and
\begin{align}\label{off1}
&\phi^s_{n,n+1}=\phi^s_{n+1,n}\nonumber\\ \nonumber
 &=\frac{(-1)^{n}\sec\mu}{16(2n+1)}
\Big[ (n+1){\sin 2n \mu}+n\sin (2n+2) \mu \Big]\cos[(2n+1)\tau\nt-\nt (n+1)\tau^s_{n+1} \nt-\nt n  \tau^s_n]\\
& \ \ \ -\frac{n(n+1)}{{8}}(\sin \mu+\mu\sec \mu)\cos[\tau\nt-\nt (n+1)\tau^s_{n+1} \nt+\nt n  \tau^s_n]\ ,
\end{align}
with $n=1,2,\cdots$.
The remaining off-diagonal contribution  $\phi^s_{m,n}\,\, (m\neq n,n\pm1)$  is given by
\begin{align}
\phi^s_{m,n} = & \frac{mn}{8\cos \mu}
\left[
\cos[n(\tau\nt-\nt\tau^s_n)\nt-\nt m(\tau\nt -\nt \tau^s_m)] \textstyle{\left( \frac{\sin(n-m+1)(\mu+\frac{\pi}{2})}{(n-m+1)(n-m)}-\frac{\sin(n-m-1)(\mu+\frac{\pi}{2})}{(n-m-1)(n-m)}\right)}
 \right.
 \nonumber\\
&\left.
-\cos[n(\tau\nt-\nt\tau^s_n)\nt+\nt m(\tau\nt -\nt \tau^s_m)] \textstyle{\left( \frac{\sin(n+m+1)(\mu+\frac{\pi}{2})}{(n+m+1)(n+m)}-\frac{\sin(n+m-1)(\mu+\frac{\pi}{2})}{(n+m-1)(n+m)}\right)}
 \right]\ ,
\end{align}
which satisfies $\phi^s_{m,n}=\phi^s_{n,m}$.
The contributions in (\ref{diag}) and (\ref{off1}) can be obtained by taking
   $\lim_{m\rightarrow n}\phi^s_{m,n}$ and $\lim_{m\rightarrow n+1}\phi^s_{m,n}$
up to some homogeneous terms.  

We note that the dilaton solution  is independent of $b_0$
since its contribution to the energy momentum tensor vanishes. Hence we shall set $b_0$ to zero for simplicity in our presentation.  Also one finds that the 
general structure of $\phi^s_{m,n}$ is similar to that of $\phi^v_{m,n}$ which is from the state deformation. Indeed one finds that the relation
\be
\phi^s_{m,n}|_{\tau^s_k=\tau^v_k+\frac{\pi}{2k}}=\phi^v_{m,n} 
\ee
works with $k=1,2, \cdots$. Hence in our study of Python's lunch geometries in Sec.~\ref{sec5}, we will 
focus on the case of the state deformation (or the source deformation alternatively) unless both 
 are turned on at the same time.

Finally one may also consider the full general deformation with 
$\chi=\chi_v+\chi_s$.
Then the corresponding dilaton solution may be obtained as
\be \label{fullsolb}
\phi = \phi_{bg} 
+8\pi G  \sum^\infty_{m=1}\sum^\infty_{n=1}\Bigl[ a_{m}a_{n} \phi^v_{m,n}+ b_{m}b_{n} \phi^s_{m,n} +2 a_{m}b_{n} \phi^c_{m,n}\Bigr]\ ,
\ee
where
\begin{align}\label{diagc}
\phi^c_{n,n} &=\frac{(-1)^{n}\, n}{8(4n^2-1)}
\left(\cos 2n \mu \tan \mu -2n{\sin 2n \mu}\right)\cos n (2\tau-\tau^v_n-\tau^s_n)\nonumber\\
&+\frac{n^2}{4}\tau \tan \mu \sin n(\tau^v_n -\tau^s_n)\ ,
\end{align}
and
\begin{align}\label{off1c}
&\phi^c_{n,n+1}=\frac{(-1)^{n+1}\sec\mu}{16(2n+1)}
\Big[ (n\nt+\nt 1){\cos 2n \mu}\nt+\nt n\cos (2n\nt+\nt 2) \mu \Big]\cos[(2n\nt+\nt 1)\tau\nt-\nt n  \tau^v_n\nt-\nt (n\nt+\nt 1)\tau^s_{n\nt+\nt 1} ]\nonumber\\
& \ \  -\frac{n(n+1)}{8}
\Bigl[ \cos \mu \cos[\tau\nt+\nt n  \tau^v_n-\nt (n\nt+\nt 1)\tau^s_{n\nt+\nt 1} \nt] -\tau \sec\mu \sin[\tau\nt+\nt n  \tau^v_n-\nt (n\nt +\nt 1)\tau^s_{n\nt +\nt 1} \nt]\Bigr]\ ,\\
&\phi^c_{n+1,n}=\frac{(-1)^{n+1}\sec\mu}{16(2n+1)}
\Big[ (n\nt+\nt 1){\cos 2n \mu}\nt+\nt n\cos (2n\nt+\nt 2) \mu \Big]\cos[(2n\nt+\nt 1)\tau\nt-\nt n  \tau^s_n\nt-\nt (n\nt+\nt 1)\tau^v_{n\nt+\nt 1} ]\nonumber\\
& \ \  +\frac{n(n+1)}{8}
\Bigl[ \cos \mu \cos[\tau\nt+\nt n  \tau^s_n-\nt (n\nt+\nt 1)\tau^v_{n\nt+\nt 1} \nt] -\tau \sec\mu \sin[\tau\nt+\nt n  \tau^s_n-\nt (n\nt+\nt1)\tau^v_{n\nt+\nt 1} \nt]\Bigr]\ ,
\end{align}
with $n=1,2,\cdots$.

The remaining term 
 $\phi^c_{m,n}\,\, (m\neq n,n\pm1)$  is given by
\begin{align}
\phi^c_{m,n} &=\frac{mn}{8\cos \mu}
\left[
\cos[\nt m(\tau\nt -\nt \tau^v_m)\nt-\nt n(\tau\nt-\nt\tau^s_n)\nt] \textstyle{\left( \frac{\cos(m-n-1)(\mu+\frac{\pi}{2})}{(m-n-1)(m-n)}-\frac{\cos(m-n+1)(\mu+\frac{\pi}{2})}{(m-n+1)(m-n)}\right)}
 \right.
 \nonumber\\
&\left.
+\cos[m(\tau\nt-\nt\tau^v_m)\nt+\nt n(\tau\nt -\nt \tau^s_n)] \textstyle{\left( \frac{\cos(m+n+1)(\mu+\frac{\pi}{2})}{(m+n+1)(m+n)}-\frac{\cos(m+n-1)(\mu+\frac{\pi}{2})}{(m+n-1)(m+n)}\right)}
 \right]\,.
\end{align}
%
Again the contributions in (\ref{diagc}) and (\ref{off1c}) can be obtained by taking
   $\lim_{m\rightarrow n}\phi^c_{m,n}$, $\lim_{m\rightarrow n}\phi^c_{m,n+1}$and $\lim_{m\rightarrow n+1}\phi^c_{m,n}$
up to some homogeneous terms. 

In the following sections, we shall analyze the asymptotic form of the above solutions 
in detail. The resulting left/right black holes are deformed in general. For instance, the Hawking temperatures of the left/right black holes become  in general different from 
each other. 
Especially when both state and source deformations are present at the same time, the asymptotic black hole spacetime is further excited rather nontrivially.

Finally one may also consider chiral deformations for which either $T_{++}=0$ and
$T_{--}=0$. (In our classical case with the massless scalar field, $T_{-+}=T_{+-}=0$ automatically.)  This type of deformations is realized by choosing $a_n =\pm\, b_n$  and
$\tau_n^s=\tau_n^v +\frac{\pi}{2n}$. Then the scalar field becomes

\begin{equation}
\label{chiralchi}
\chi_\pm = \sum^\infty_{n=1} b_n \sin \big(\tau \pm \mu \pm \frac{\pi}{2}-\tau^v_n \big) \,,
\end{equation}
which gives us  $T_{\pm\pm}=(\partial_\pm\chi_\pm)^2$ with $T_{\mp\mp}=0$, respectively. Here,
we introduced $x^\pm =
\tau \pm \mu$. In these coordinates, the equation of motion for the dilaton 
involving $T_{\mp\mp}$ becomes
\begin{equation} \label{eomnull}
-\sec^2\mu\, \partial_{\mp} (\cos^{2}\mu\, \partial_{\mp}\phi ) = T_{\mp\mp} \ .
\end{equation}
Thus one may show that 
the dilaton field can be written in the form of
\begin{equation} 
 \label{chiralphi}
\phi =f_{\pm} (
x^\pm) \tan \mu + g_{\pm} (
x^\pm) \ ,
\end{equation}
for  $\chi=\chi_\pm(x^\pm)$, respectively.

\section{Deformations of two-sided black holes  
}\label{sec4}
In this section, we shall analyze the resulting deformation of the two-sided  
black hole spacetime including the horizon structure and the late-time 
temperature of the black hole system. 

Physical properties of the deformed system can be obtained 
by studying the asymptotic behavior of the dilaton solution. 
We shall see that, when only one kind of deformation of the two in 
\eqref{fulldefor} is turned on, the asymptotic behavior of the 
dilaton reduces to the form of the vacuum solution \eqref{dilaton} with
deformed constants. In general the Hawking temperatures of the 
left/right black holes changes differently from each other. Although
there can be many nontrivial Python's lunches developed in the bulk, 
the boundary dynamics will be that of simply deformed black holes with 
corresponding modifications in the cut-off trajectory and does not directly capture
the effect of the Python's lunches. In the presence of both the state
deformation and the source deformation, the asymptotic behavior
of the dilaton turns out to be more complicated than the vacuum solution
\eqref{dilaton}. This would be then reflected in the boundary dynamics as
a new term in the boundary action as well as more complicated trajectories.

To see first the causal structure of the solution (\ref{fullsol}), let us  study its asymptotic behavior as $\mu \rightarrow \pm\frac{\pi}{2}$. Note that, 
as $\mu \rightarrow \pm\frac{\pi}{2}$,
\begin{align}
\phi^v_{n,n}\  &=
-\frac{n^2}{4}(1+\mu\tan \mu) +{\cal O}(\cos^2 \mu)\ ,\nonumber\\
\phi^v_{n,n+1}&=\phi^v_{n+1,n}= -\frac{n(n+1)}{8}(\sin \mu+\mu\sec
\mu)\cos[\tau\nt-\nt (n+1)\tau^v_{n+1} \nt+\nt n  \tau^v_n]+{\cal O}(\cos^2 \mu)\ ,\nonumber
\end{align}
while all the remaining of $\phi^v_{m,n}$ are ${\cal O}(\cos^2 \mu)$. We shall also 
introduce $\alpha_n$ by the rescaling 
$\alpha_n = n a_n \sqrt{\frac{\pi^2 G}{\bar\phi L_0}}$. 
The dilaton in asymptotic regions  can then be written in the form of  \eqref{phias}
with
\be
Q(\tau)=\cos \tau -\sum^\infty_{n=1} \alpha_n^2 +(\pm) \sum^\infty_{n=1} \alpha_n
\alpha_{n+1}\cos[\tau\nt-\nt (n+1)\tau^v_{n+1} \nt+\nt n  \tau^v_n]\ .
\ee
Here and below the upper/lower sign inside parenthesis is  for the left/right boundary region respectively. 

\begin{figure}
\begin{center}
\includegraphics[width=6.5cm,clip]{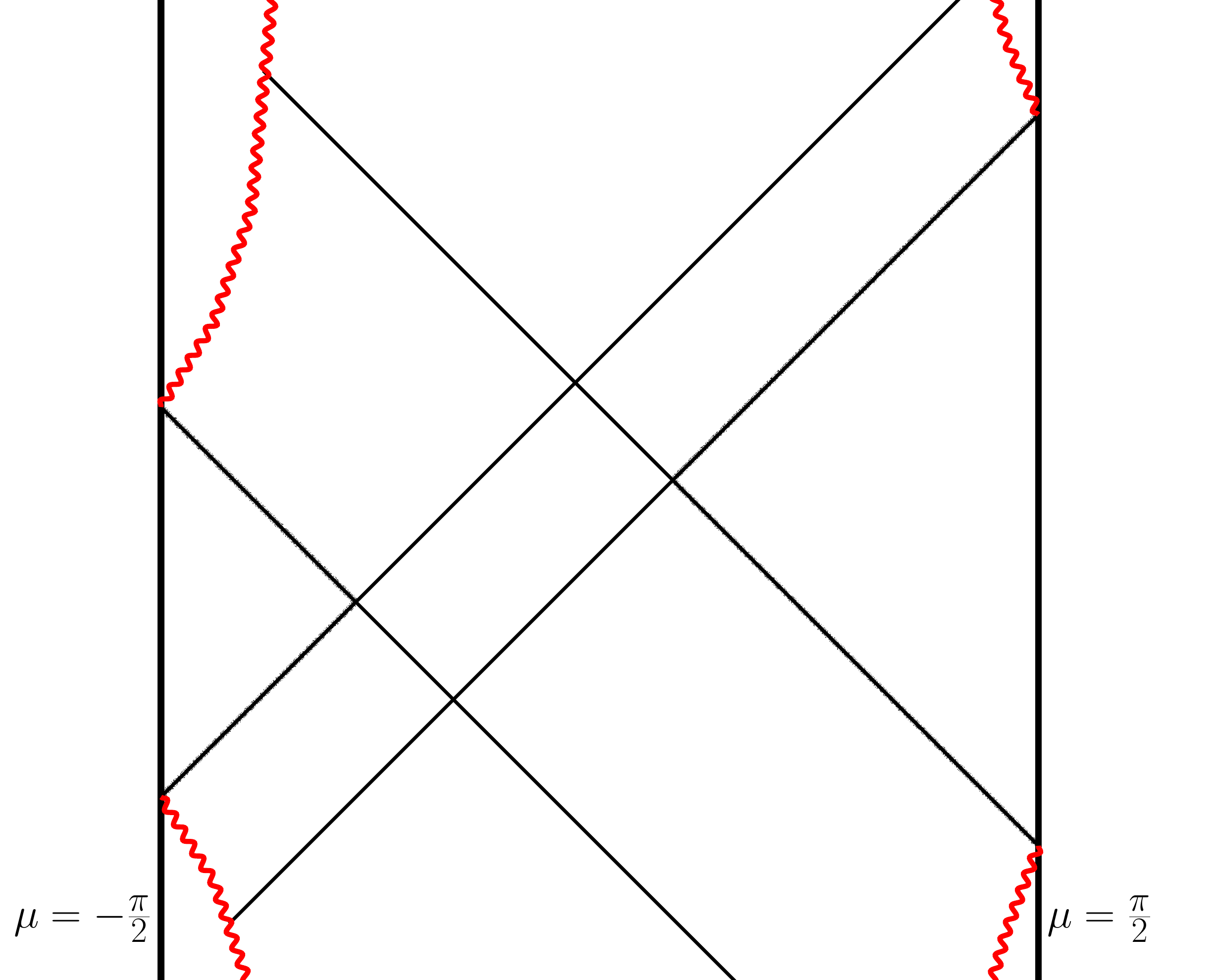}
\end{center}
\vskip-0.5cm
\caption{Penrose diagram of deformed two-sided black hole spacetime.}
\label{deformed} 
\end{figure}

 As was noted previously~\cite{Maldacena:2016upp}, the reparameterization can be fixed by the prescription of the dilaton cutoff
\begin{equation} \label{}
\phi|_{\rm cutoff} =\bar\phi \frac{\ell}{\epsilon}\ ,
\end{equation}
treating $\phi$ as a radial coordinate in the asymptotic region. This will fix the asymptotic left/right cutoff trajectory $\mu^{L/R}_c (\tau)$. For simplicity below, we shall frequently drop the index $L$ and $R$ since the left and right asymptotic regions can be treated in the same way. Through the asymptotic form of the dilaton field in~\eqref{phias}, the trajectory is then given by\be
\cos \mu_c =\frac{L_0}{\ell} Q(\tau) \epsilon\ .
\ee
The reparameterization on this trajectory enables us to express $\tau$ in terms of the boundary time ${u}$. Namely, from the prescription in the metric in~\eqref{Gmetric}, we have
\be
ds^2|_{\rm cutoff}\left(=-\frac{\ell^2 d \tau^2}{\cos^2 \mu_c}\right)
 = -\frac{du^2}{\epsilon^2}\ .
\ee  
This leads to
\be
du =d\tau \frac{\ell^2}{L_0 Q(\tau)}\ .
\ee
We arrange $Q(\tau)$ as in the form of
\be
Q(\tau)=\sqrt{A^2+B^2}(\cos (\tau-\tau_B) -q)\ ,
\ee
where
\begin{align} 
A \, &= 1+(\pm) \sum^\infty_{n=1} \alpha_n \alpha_{n+1}\cos[\nt (n+1)\tau^v_{n+1}
\nt-\nt n  \tau^v_n]\ ,
\nonumber\\
B \, &= (\pm) \sum^\infty_{n=1} \alpha_n \alpha_{n+1}\sin[\nt (n+1)\tau^v_{n+1}
\nt-\nt n  \tau^v_n]\ ,
\nonumber\\
q \ &= \frac{\sum^\infty_{n=1} \alpha^2_n}{\sqrt{A^2+B^2}}\ ,\nonumber\\
 \tau_B &= \arctan \frac{B}{A}.    \label{horizon}
\end{align}
For the asymptotic region to be well-defined, $q <1$ and $A>0$ is required. Then the reparameterization is
solved by
\begin{equation} \label{Relutau}
\tanh \frac{L}{\ell^2}\frac{u}{2} =\sqrt{\frac{1+q}{1-q}} \tan \frac{\tau-\tau_B}{2}\ ,
\end{equation}
where $L= L_0 \sqrt{A^2+B^2}\sqrt{1-q^2}$. Since $u$ is the boundary time coordinate of the cutoff trajectory in the Rindler wedge in 
(\ref{btz}), the late-time temperature of
the deformed system becomes
\be
T=T_0\sqrt{A^2+B^2}\sqrt{1-q^2}\ .
\ee
One further find that the future/past infinity of the Rindler wedge is at
\be \label{matching}
\tau_{f/p}=  \pm 2 \arctan \sqrt{\frac{1-q}{1+q}}+ \tau_B \ ,
\ee
which can also be obtained from $Q(\tau)=0$.
Then the L/R future horizons
are described by straight lines  
\begin{equation} \label{}
(\mp)\mu=(\tau-\tau^{L/R}_f) +\frac{\pi}{2}\ ,
\end{equation}
while the L/R past horizons are by
\be
(\pm)\mu=(\tau-\tau^{L/R}_p) -\frac{\pi}{2}\ .
\ee
The corresponding Penrose diagram is depicted in Figure \ref{deformed}.
Note that in this figure the end points of the future/past horizon 
$\tau_{f/p}$ are shifted from $\pm \frac\pi2$ to values with 
$|\tau_{f/p}| \,      \le\, \frac\pi2$. 
This can be seen from
%
\be \label{costau}
\cos\tau_{f/p} = \frac1{A^2 + B^2}
  [ A \alpha \mp {\rm sign}(B)\sqrt{A^2\alpha^2 - (A^2 + B^2) (\alpha^2 - B^2)} ]\ ,
\ee
where $\alpha \equiv \sum^\infty_{n=1} \alpha^2_n $. 
Since 
\begin{align}
\alpha \pm B 
&= \frac12 \alpha_1^2 +\frac12 \sum^\infty_{n=1} \left\{ \alpha_n^2
+ \alpha_{n+1}^2 \pm(\pm) 2 \alpha_n \alpha_{n+1} \sin[(n+1)\tau^v_{n+1} - n
\tau^v_n]
\right\} \ ,\nonumber \\
&\ge    0\ ,
\end{align}
the numerator of \eqref{costau} is positive semidefinite, 
which proves $|\tau_{f/p}| \, \le \,    \frac\pi2$.

In case of the source deformations where the dilaton is given by 
\eqref{fullsolb}, the functional form of each mode $\phi_{m,n}^s$ is 
essentially the same as that of the state deformation $\phi^v_{m,n}$ and hence
we get the same results. However, for the full general deformations 
\eqref{fulldefor} including both state as well as source deformations, 
we obtain rather different behaviors due to the cross terms $\phi_{m,n}^c$ in
\eqref{fullsolb}. To see this, let us first consider the simplest case
where both the state and the source deformations are turned on,
\be
\chi=a_1 \cos \mu \cos (\tau-\tau^v_1) - b_1 \sin \mu \cos (\tau-\tau^s_1) \ ,
\ee 
with the corresponding dilation solution 
\be
\phi = \phi_{bg} 
+8\pi G ( a_1^2\phi^v_{1,1}+ b_1^2 \phi^s_{1,1} +2 a_1 b_1 \phi^c_{1,1} )\ .
\ee
As $\mu \rightarrow (\mp) \frac\pi2$, $\phi$ behaves as in \eqref{phias} with
\be
Q(\tau) = \cos\tau - \alpha_1^2 - \beta_1^2 
 +(\mp) 2 \alpha_1 \beta_1 \left[ \frac1{3\pi} \cos(2\tau-\tau^v_1-\tau_1^s)
			     + \frac{2\tau}\pi \sin(\tau^v_1-\tau_1^s) \right]\ ,
\ee
where $\beta_n = n b_n \sqrt{\frac{\pi^2 G}{\bar\phi L_0}}$. Note that
the cross term in $Q(\tau)$ contains a linear term in $\tau$ which cannot
be obtained as a function of $u$ in a closed form. For small deformations,
one can still compute future/past infinities at which $Q(\tau)$ vanishes.
The results are
\be
\tau_{f/p} = \pm\frac\pi2 + \delta \tau_{f/p}\ ,
\ee
where
\begin{align} \label{deltataufp}
\delta \tau_f
 &= -\alpha_1^2 - \beta_1^2 +(\mp) 2 \alpha_1 \beta_1 \left[ 
   \sin(\tau^v_1 - \tau_1^s) - \frac1{3\pi} \cos(\tau^v_1 + \tau_1^s) \right]
	+ \cdots \ ,\nonumber \\
\delta \tau_p
 &= \alpha_1^2 + \beta_1^2 +(\mp) 2 \alpha_1 \beta_1 \left[
   \sin(\tau^v_1 - \tau_1^s) + \frac1{3\pi} \cos(\tau^v_1 + \tau_1^s) \right]
	+ \cdots\ .
\end{align}
Here the ellipses are quartic order terms in $\alpha_1$ and $\beta_1$
and the upper/lower sign refers to the left/right boundary respectively as before.
Note that $\delta \tau_{f/p}$ is not negative/positive definite. 
In other words, $|\tau_{f/p}|$ can be larger than $\frac\pi2$ 
when both the state and the source are deformed together, 
in contrast with the previous cases where only one kind of deformation is 
turned on. Nevertheless,  for this particular deformation with
the lowest modes,  the difference of the future infinity and the past 
infinity is still less than or equal to  $\pi$,  {\it i.e.},
\begin{align} \label{tauftaup}
\tau^{L/R}_f - \tau^{L/R}_p &= \pi - 2(\alpha_1^2 + \beta_1^2)
+ (\pm) \frac4{3\pi} \alpha_1 \beta_1 \cos(\tau^v_1 + \tau_1^s) + \cdots\nonumber\\
	&\le \pi\ .
\end{align}
Then the difference becomes smaller by the perturbation,
although the overall spacetime is slightly shifted either upward or downward.

From \eqref{deltataufp}, one can also compare shifted values in other
boundaries.
Suppose for instance that $\delta\tau_f^{R} \ge 0$.  (The other case $\delta\tau_f^{L} \ge 0$ can be treated in a similar manner.)
Assuming $\alpha_1 \beta_1 \ge 0$,
this happens only when
\be
\sin(\tau^v_1 - \tau_1^2) - \frac1{3\pi} \cos (\tau^v_1 + \tau_1^s) \ge 1\ .
\ee
Then it is not difficult to see that the following relation holds
\be
\delta\tau_f^L \le 0 \le \delta\tau_f^R \le \delta\tau_p^L 
\le \delta\tau_p^R \ . 
\ee
Therefore although the position $\tau_f^R$ of the future infinity at the
right boundary is shifted slightly upward, the other three positions are
also moved enough to compensate for such shift. 
In particular, $\delta\tau_f^{L/R} \le \delta\tau_p^{R/L}$ implies that the causal 
structure is not of the type of traversable wormholes, as it should be for 
classical perturbations.

Having seen what happens when both the state and the source deformations are
turned on, let us now consider the most general deformation \eqref{fulldefor}.
Denote
\be
Q(\tau) = \cos\tau + Q_1^{L/R}(\tau)\ ,
\ee
where $Q_1^{L/R}(\tau)$ is given by
\be
Q_1^{L/R} = \sum_{m,n} ( \alpha_m \alpha_n \varphi^{v,L/R}_{m,n} 
       + \beta_m \beta_n \varphi_{m,n}^{s,L/R}
       + 2 \alpha_m \beta_n \varphi_{m,n}^{c,L/R} )\ ,
\ee
with 
\be
\varphi_{m,n}^{L/R}(\tau) \equiv \left. \frac{8\cos\mu}{\pi m n} 
           \phi_{m,n}(\tau,\mu)\right|_{\mu \rightarrow \mp \frac\pi2}\ .
\ee
Then
\be
\tau_f^R - \tau_p^R = \pi + Q_1^R(\tfrac\pi2)
		+ Q_1^R(-\tfrac\pi2) + \cdots\ ,
\ee
where ellipsis is of quartic or higher order in $\alpha_n$'s and $\beta_n$'s.
Nonvanishing contributions from the cross term $\varphi_{n,m}^{c,R}$ are
\begin{align}
\varphi_{n,n}^{c,R}(\tfrac\pi2) +\varphi_{n,n}^{c,R}(-\tfrac\pi2)
    \  \ &= \frac{2(-1)^n}{\pi n (4n^2 - 1)} \cos[n(\tau^v_n + \tau_n^s)] ,
                                                     \nonumber \\
\varphi_{n,n+1}^{c,R}(\tfrac\pi2) +\varphi_{n,n+1}^{c,R}(-\tfrac\pi2)
     &= \cos[n\tau^v_n -(n+1) \tau_{n+1}^s], \nonumber \\
\varphi_{n+1,n}^{c,R}(\tfrac\pi2) +\varphi_{n+1,n}^{c,R}(-\tfrac\pi2)
     &= -\cos[n\tau^s_n -(n+1) \tau_{n+1}^v], 
\end{align}
and for $m\neq n$ and $m-n$ even,
{\small
\be \label{phicmn2}
\varphi_{m,n}^{c,R}(\tfrac\pi2) + \varphi_{m,n}^{c,R}(-\tfrac\pi2)
  = (- 1)^{\frac{m-n}2+1} \left\{ \negthinspace \negthinspace
       \frac{4\cos(m \tau^v_m - n \tau_n^s)}{\pi(m\negthinspace-\negthinspace n)[(m\negthinspace-\negthinspace n)^2\negthinspace -\negthinspace 1]}
     - \frac{4\cos(m \tau^v_m + n \tau_n^s)}{\pi(m\negthinspace+\negthinspace n)[(m\negthinspace+\negthinspace n)^2 \negthinspace-\negthinspace 1]} \right\}.
\ee
}
Collecting terms from $\varphi^v_{m,n}$, $\varphi_{m,n}^s$ and  $\varphi_{m,n}^c$, we have
\begin{align} \label{q1q1}
&Q_1^R(\tfrac\pi2) +  Q_1^R(-\tfrac\pi2) \\
 = &-2\sum_n \left\{ \alpha_n^2 + \beta_n^2 \right\}+ 2\hskip-0.3cm \sum_{m-n\ \textrm{even}} \hskip-0.3cm \alpha_m \beta_n \,
   [\varphi_{m,n}^{c,R}(\tfrac\pi2) + \varphi_{m,n}^{c,R}(-\tfrac\pi2)]\ 
\nonumber \\
&+2\sum_n \left\{
 \alpha_n \beta_{n+1} [\varphi_{n,n+1}^{c,R}(\tfrac\pi2) + \varphi_{n,n+1}^{c,R}(-\tfrac\pi2)] 
+ \alpha_{n+1} \beta_{n} [\varphi_{n+1,n}^{c,R}(\tfrac\pi2) + \varphi_{n+1,n}^{c,R}(-\tfrac\pi2)] 
\right\}\,. \nonumber
\end{align}
We can explicitly diagonalize \eqref{q1q1} for some finite number of 
deformations. If $\alpha_n=\beta_n=0$ for $n \ge 8$, we find that 
\eqref{q1q1} is negative semidefinite. If higher modes are present, however,
it turns out that \eqref{q1q1} can be positive because of the cross terms.
In other words, if high 
enough modes of both state and source deformations are turned on at the 
same time, the difference between the future infinity and the past infinity 
in one side can be larger than $\pi$.
Since $|\tau_{f/p}| \le \frac\pi2$ in the presence of only one type of 
deformations as seen above, this result is due to the 
interaction of both deformations for which the asymptotic black hole spacetime
is nontrivially excited from the vacuum solution. Also, the trajectories of 
the boundary dynamics would be deformed accordingly.

On the other hand, the difference of the future infinity of one side
and the past infinity of the other side must still be less than or equal
to $\pi$ for arbitrary deformations, since traversable wormholes should not 
be formed. Indeed, we find that
\begin{equation}
\tau_f^R - \tau_p^L = \pi + Q_1^R(\tfrac\pi2) + Q_1^L(-\tfrac\pi2) + \cdots\ ,
\end{equation}
where $Q_1^R(\tfrac\pi2) + Q_1^L(-\tfrac\pi2)$ can be written in
a manifestly negative semidefinite form,
\begin{align}
Q_1^R&(\tfrac\pi2) + Q_1^L(-\tfrac\pi2)  \\
 = &-(\alpha_1 \sin \tau^v_1 -\beta_1 \cos \tau^s_1)^2
    -(\alpha_1 \cos \tau^v_1 +\beta_1 \sin \tau^s_1)^2 \nonumber \\
 & -\sum_n \left\{ [\alpha_n \sin n\tau^v_n -\beta_n \cos n\tau^s_n
    -\alpha_{n+1} \cos (n+1)\tau^v_{n+1}
    -\beta_{n+1} \sin (n+1)\tau^s_{n+1} ]^2 \right. \nonumber \\
 & \phantom{-\sum_n\ }\left. +[\alpha_{n+1} \sin (n+1)\tau^v_{n+1}
    - \beta_{n+1} \cos (n+1)\tau^s_{n+1}
    + \alpha_n \cos n\tau^v_n +\beta_n \sin n\tau^s_n]^2 \right\}. \nonumber
\end{align}

Since the general dilaton solution \eqref{fullsolb} is an infinite series
of trigonometric functions with arbitrary coefficients, one may wonder
if the extremum points of $\phi$ can exist anywhere in the spacetime. We
shall argue, however, that classically they should reside behind the horizon
thanks to the NEC. Let us focus on the right Rindler wedge
for definiteness. In terms of the affine parameter for a fixed 
$x^- = x^-_0$,
\begin{equation} \label{}
\lambda = \tan\left( \frac{x^{+}-x^{-}_{0}}2 \right),
\end{equation}
the equation of motion \eqref{eomnull} in null coordinates
can be written as
\begin{equation} \label{}
\frac{d^{2}\phi }{d\lambda^{2}}
 = -4\cos^{4}\left( \frac{x^{+}-x^{-}_{0}}2 \right)~ T_{++}
 = -T_{\lambda\lambda}\,,
\end{equation}
which is the two-dimensional analogue of the Raychaudhuri equation.  Here, $\frac{d}{d\lambda}\phi$ plays the role of the expansion of the null geodesic congruence. 
Now, the positive NEC ($T_{\lambda\lambda} >  0$) implies
\begin{equation} \label{}
\frac{d^{2}\phi }{d\lambda^{2}} <  0\,,
\end{equation}
which shows us that null rays from the extremal point of $\phi$,  defined 
by $\partial_{+}\phi =\partial_{-}\phi =0$,  should be met with a 
singularity within a finite affine parameter and so the extremal points should reside behind the horizon. Note that this is nothing 
but a two-dimensional version of the famous singularity theorem or the 
focusing argument~\cite{Hawking:1973uf,Bousso:2015mna,Engelhardt:2021qjs}. 

\begin{figure}
\begin{center}
\includegraphics[width=6cm,clip]{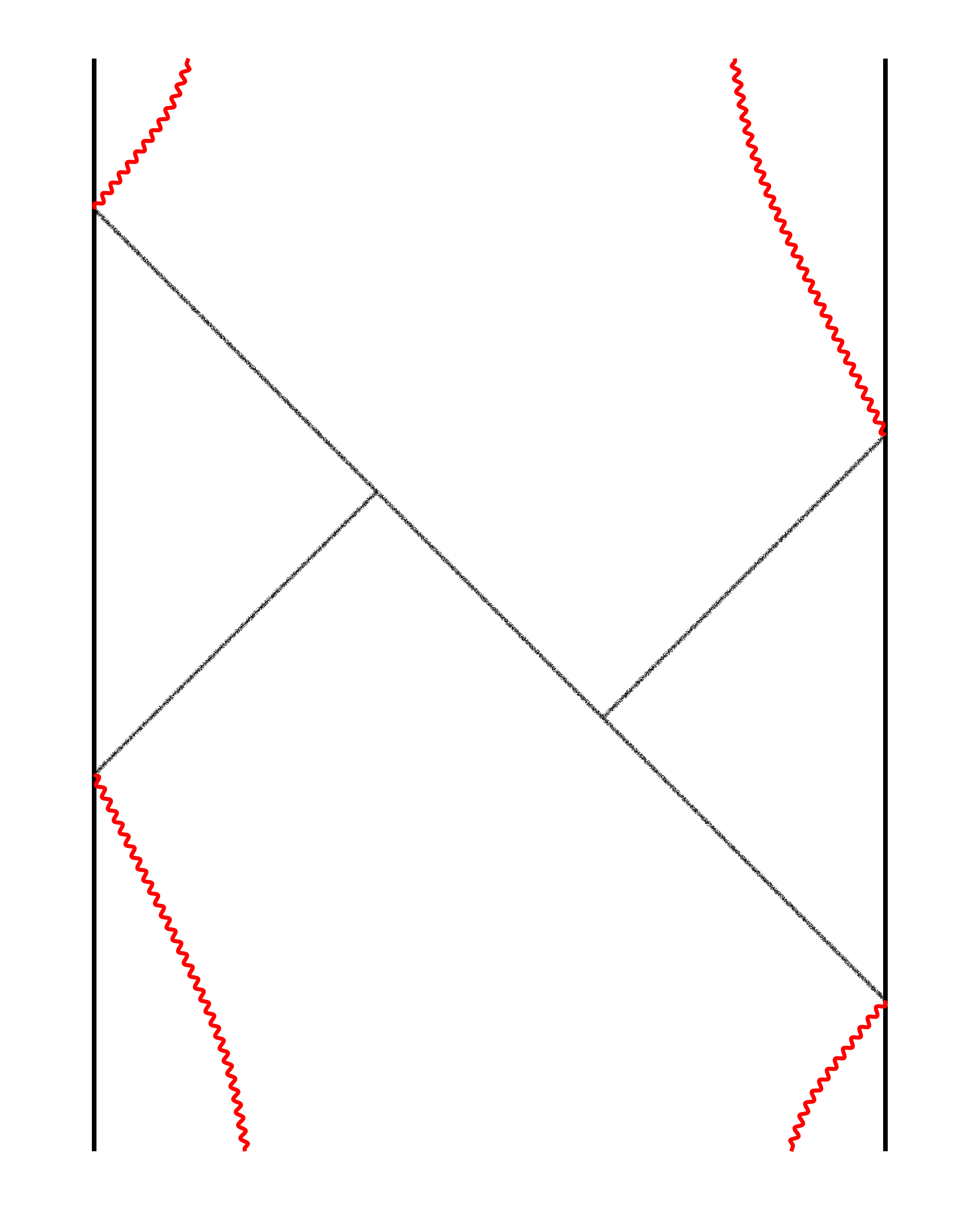}
\end{center}
\vskip-0.5cm
\caption{Penrose diagram of the two-sided black hole 
	spacetime deformed by the antichiral matter $\chi=\chi(x^+)$.}
\label{chiralhorizon} 
\end{figure}

In case of the chiral deformations \eqref{chiralchi}, the situation is
rather simple. For definiteness, 
let us choose the upper sign in \eqref{chiralchi} which corresponds to 
$T_{--}=0$. Then there are only ingoing/outgoing modes in the right/left
sides and we expect that the future/past horizons of the right/left black holes
would be deformed to outwards/inwards so that they do not meet at one point.
On the other hand, the other side of the horizons, namely the past/future 
horizons of the right/left black holes would be still connected as a single
straight line. See Figure~\ref{chiralhorizon}. 
This can easily be seen from the general form of the dilaton 
solution \eqref{chiralphi}. Note that at the horizon we would have 
$\frac{d\phi}{d\lambda}\Big|_{H} \underset{\lambda= \pm\infty}=0$, where
$\lambda$ is the relevant affine parameter. It 
implies the equation $f_+(x^+)=0$ which is independent of $x^-$. 
Then in the spacetime under consideration where the horizon is unique at each
asymptotic region, the solution $x^+ = x^+_H$ should define a unique
straight-line throughout the spacetime. In other words, it represents 
the past/future horizons on the right/left sides connected as a single 
straight line as expected. 

We now show that the extremal point should be unique and lie on the 
horizon for chiral deformations. 
First, it is evident that $\partial_-\phi = 0$ results
in $f_+(x^+)=0$ which is nothing but the defining equation of the horizon
$x^+=x^+_H$. The other condition $\partial_+ \phi = 0$ for the
extremal point gives
\begin{equation}
	\partial_+\phi = f_+'(x^+) \tan\mu + g_+'(x^+) = 0\ ,
\end{equation}
which has a unique solution on the horizon $x^+ = x^+_H$,
\begin{equation}
	x^- = x^+_H + 2 \tan^{-1}\left(\frac{g_+'(x^+_H)}{f_+'(x^+_H)}\right)\ ,
\end{equation}
which completes the proof. Note in particular that there cannot be a
Python's lunch at least classically for chiral deformations.


\section{Python's lunch geometries  
}\label{sec5}

\begin{figure}[htbp]  
\centering
\includegraphics[width=0.4\textwidth]{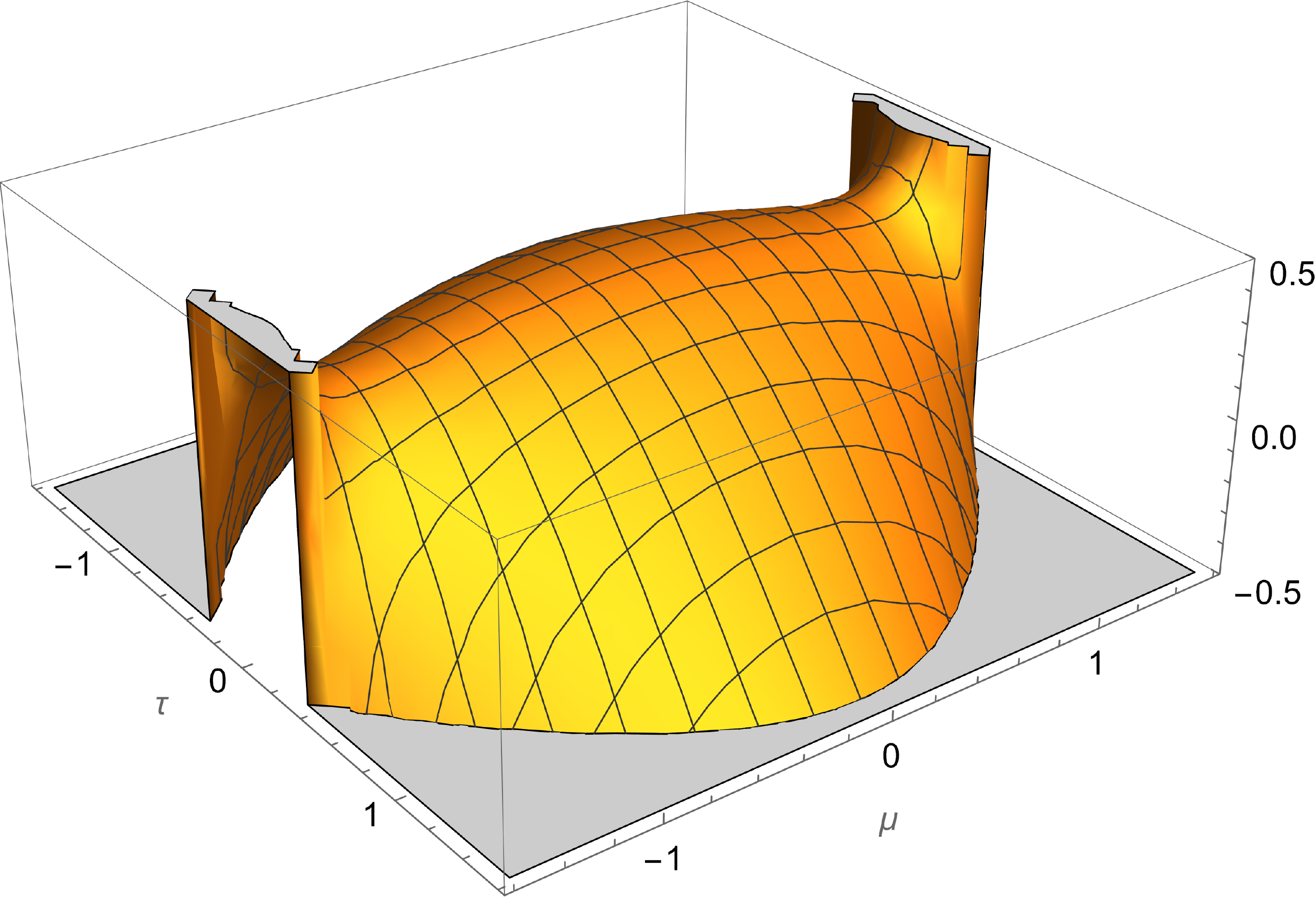}
\caption{3D plot of dilaton field for Janus deformation with $\gamma^2=6/5$.}
 \label{pfigj}
\end{figure}

In this section, we discuss Python's lunch geometries formed by the
general deformations. Let us first consider the 
Janus deformation \eqref{janusdef} to identify a Python's lunch in the 
simplest setup. Since $\cos\tau$ is the only $\tau$-dependent term in
the dilaton configuration \eqref{janusdef},
we should consider the $\tau=0$ slice to find any extremum point. 
At $\tau=0$, the dilaton field can be written as
\begin{equation}
	\phi = \bar{\phi}L_{0}\left[\frac1{\cos\mu}
	       -\frac{\gamma^2}2 (1+\mu\tan\mu)\right]\,,
\end{equation}
where we set $\frac{8\pi G}{\bar\phi L_0} \equiv 1 $ in this section.
Then as $\mu\rightarrow \frac\pi2$,
\begin{equation}
\phi = \frac{1-\frac\pi4 \gamma^2}{\frac\pi2 - \mu}
            + O\left(\frac\pi2 - \mu \right),
\end{equation}
which requires $\gamma^2 <\frac4\pi$.
For $\mu \simeq 0$, $\phi$ becomes
\begin{equation}
\phi = \bar{\phi}L_{0}\left[\frac12(1-\delta) - \frac\delta2 \mu^2
	+\frac1{24} ( 1 - 4\delta ) \mu^4 \right] + O(\mu^6),
\end{equation}
where $ \delta = \gamma^2 - 1$.
It is clear that, for small positive $\delta$, $\mu=0$ is a maximum point 
of $\phi$. The QES$_{\rm min}$ is the minimum point of $\phi$ which occurs at 
$\mu_q \simeq \sqrt{6\delta}(1+O(\delta))$ with
\begin{equation}
\phi(\mu_q) = \bar{\phi}L_{0}\left[
	\frac12(1-\delta) - \frac32\delta^2 \right] + O(\delta^3).
\end{equation}
Therefore, for $1 < \gamma^2 < \frac4\pi$,
a bulge is formed at $\mu=0$ and QES$_{\rm min}$ is located at 
$\mu_q \simeq \sqrt{6\delta}$. The difference of the two dilaton values is
\begin{equation}
\phi(0) - \phi(\mu_q)  = \frac32 \bar\phi L_0 \delta^2 + O(\delta^3).
\end{equation}
We draw the dilaton field under the Janus deformation in Figure~\ref{pfigj}.

\begin{figure}[tbph]
\vskip-0.5cm   
\centering
\subfigure[]
{ \label{fig:pfiga}
   \includegraphics[width=0.36\textwidth]{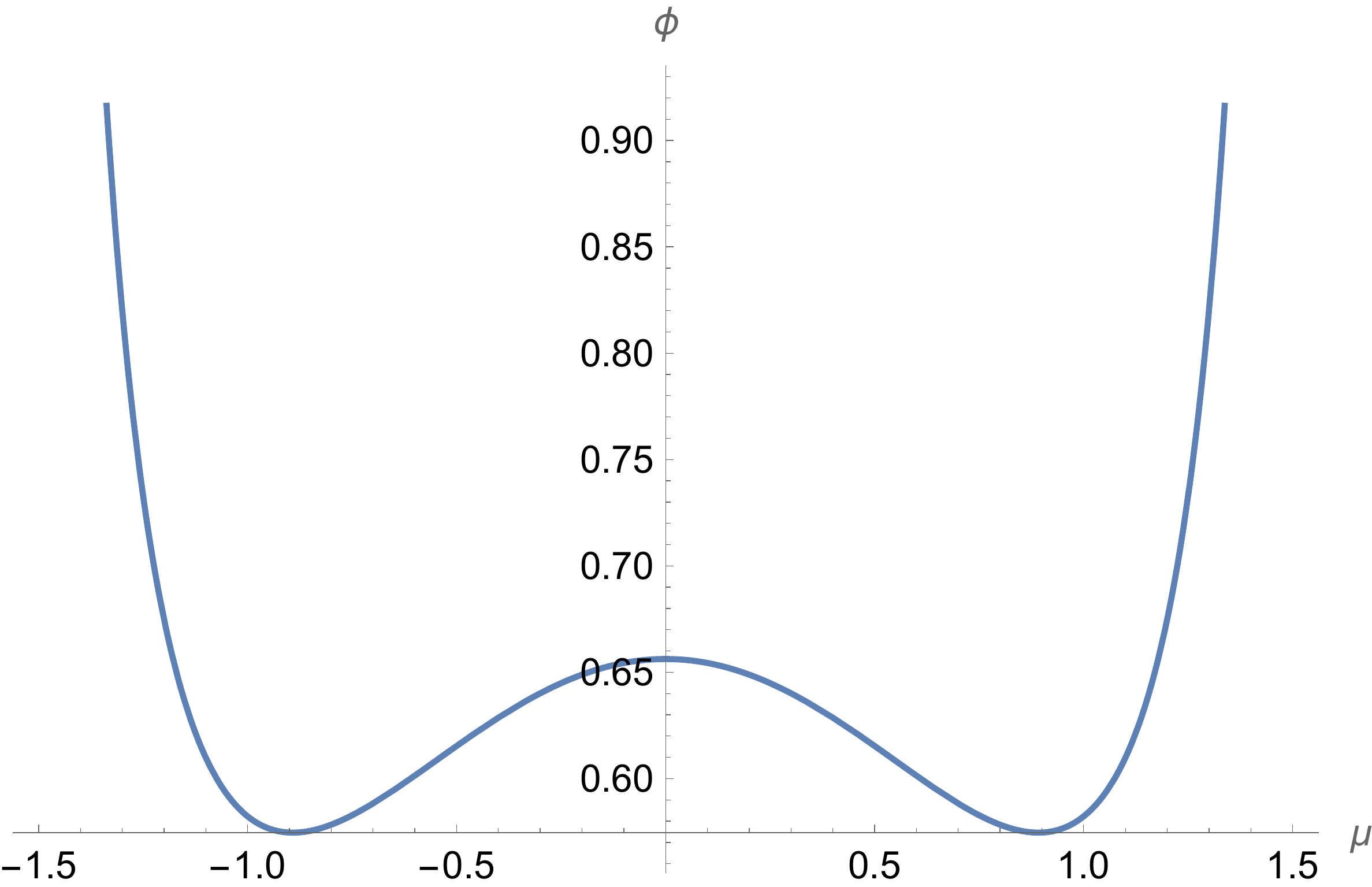}
}
\subfigure[]
{ \label{fig:pfigb}
   \includegraphics[width=0.36\textwidth]{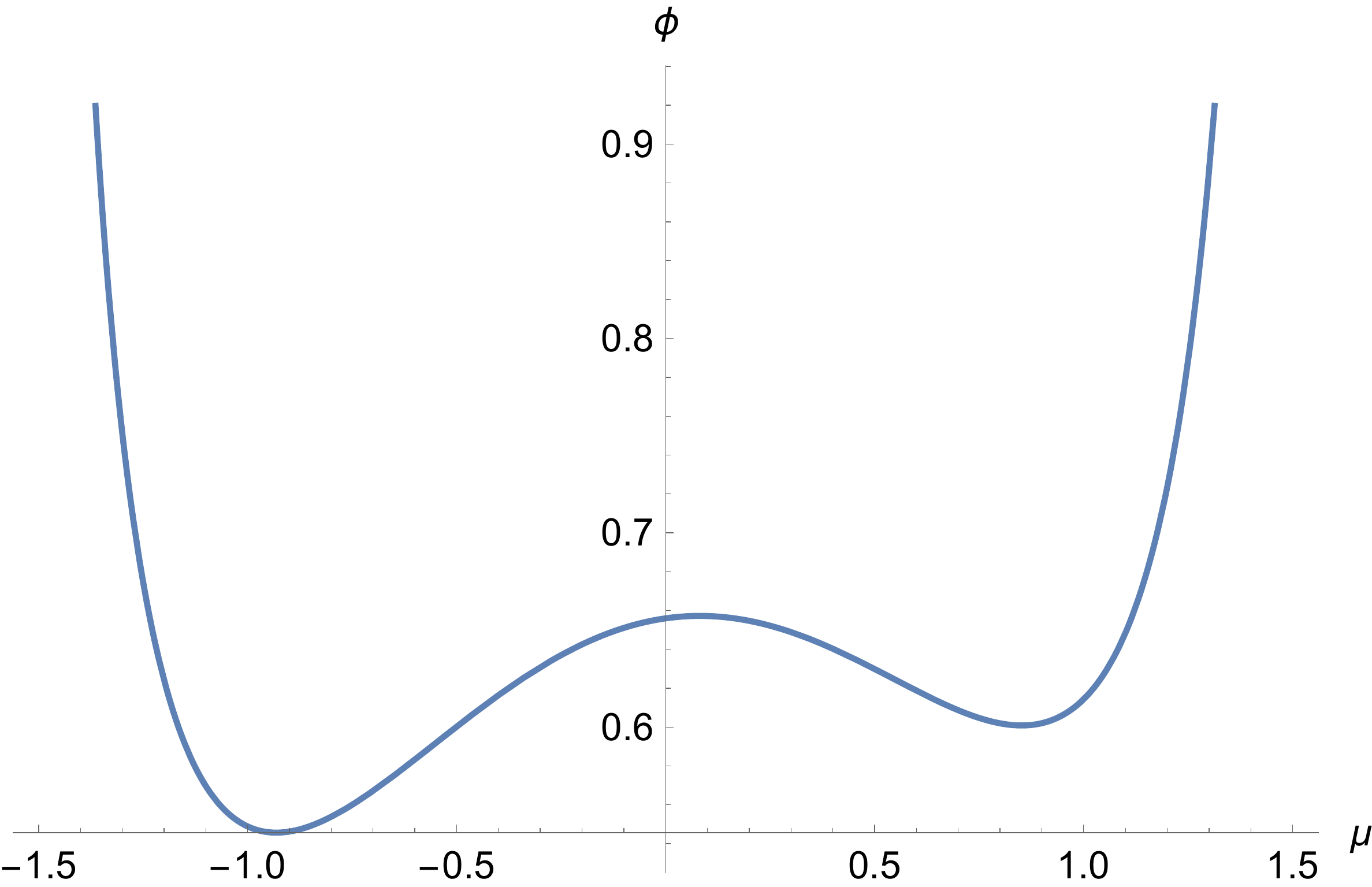}
}
\subfigure[]
{ \label{fig:pfigc}
   \includegraphics[width=0.36\textwidth]{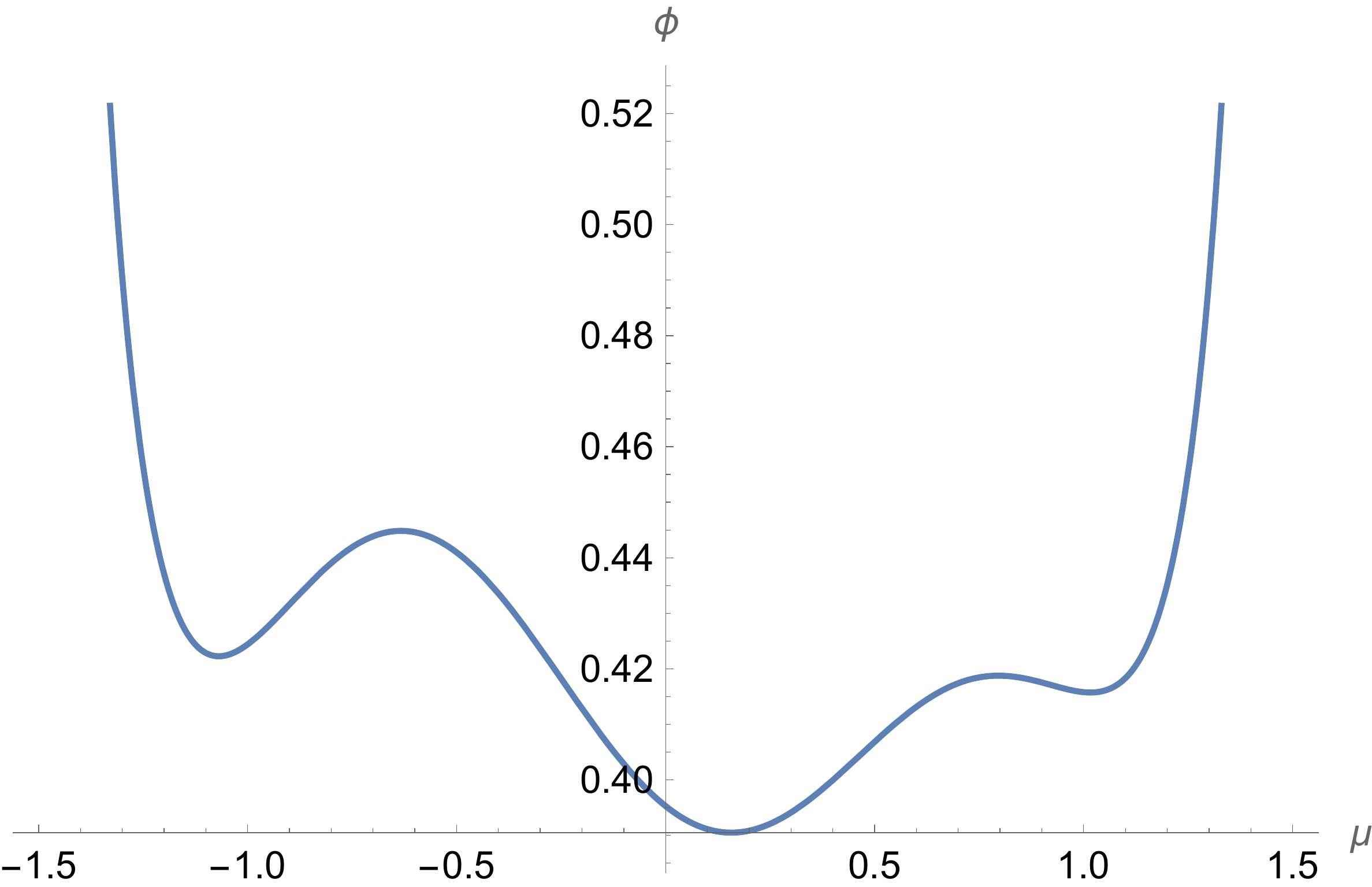}
}
\subfigure[]
{ \label{fig:pfigd}
   \includegraphics[width=0.36\textwidth]{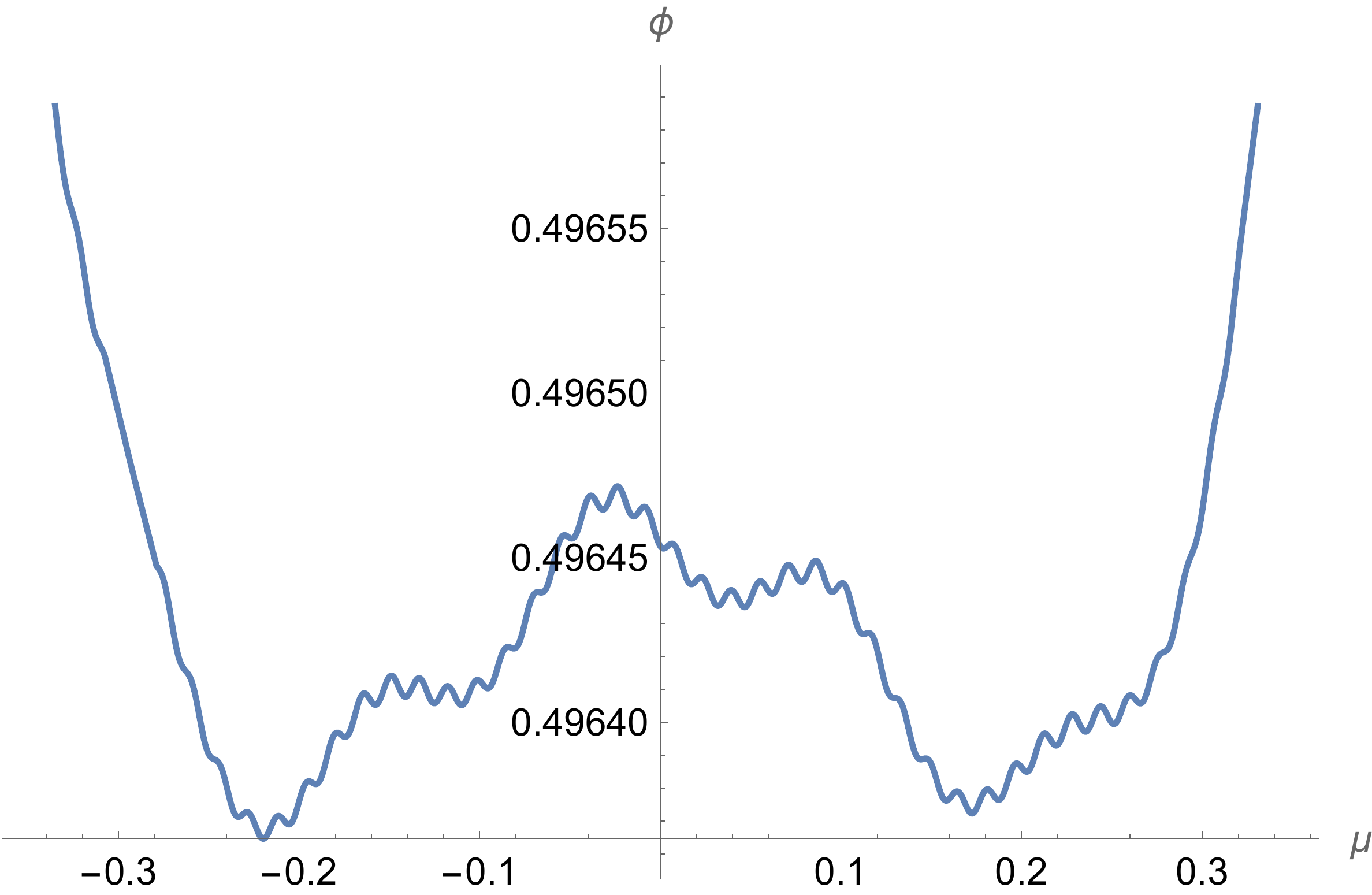}
}
\vskip-0.3cm
\caption{\small Various plots of dilaton field with $\tau=0$ and $\tau_n = \frac\pi{2n}$.
(a) For $\chi = a_1 \chi_1^v$ with $a_1 = 0.9\sqrt{
8/\pi}$.
(b) For $\chi = a_1 \chi_1^v + a_2 \chi_2^v$
with $a_1 = 0.9\sqrt{8/\pi}$ and $a_2 = -0.02\sqrt{8/\pi}$.
(c) For $\chi = a_1 \chi_1^v + a_2 \chi_2^v + a_3 \chi_3^v$
with $a_1 = 0.094\sqrt{8/\pi}$, $a_2 = 0.94\sqrt{8/\pi}$ and 
$a_3 = -0.094\sqrt{8/\pi}$ .
(d) For $\chi = a_{151} \chi_{151}^v + a_{200} \chi_{200}^v$
with $a_1 = 0.0316\sqrt{8/\pi}$, $a_2^2 = 0.8888\sqrt{8/\pi}$.
}
\end{figure}

Now we look into more general deformations analyzed in section~\ref{sec3}.
For the lowest-mode state deformation $ \chi = a_1 \chi_1^v $, the dilaton
solution is given by \eqref{pv123} and \eqref{phiv123}.
Choosing $\tau_1^v = \frac\pi2$ and $a_1$ such that 
$\frac32 < a_1^2 < \frac8\pi$, the dilaton is positive definite 
and has a maximum at $\mu=\tau=0$ and hence a Python's lunch geometry
is formed around the origin, as depicted in Figure~\ref{fig:pfiga}. 
For $a_1^2 \simeq 3/2$, the QES$_{\rm min}$ occurs at $\mu \simeq 0$ while, 
as $a_1$ increases, its position moves towards $\mu = \frac\pi2$.

So far, we have considered left-right symmetric configurations. The QES$_{\rm min}$ in
this case is located at two places, one in $\mu>0$ region and
the other in $\mu<0$ region, respectively, since the dilaton fields are
even functions of $\mu$. If we turn on several modes at the same time,
the dilaton has cross terms which can be odd in $\mu$ as seen in
\eqref{off1n}. Then the degeneracy is broken and we have only one QES$_{\rm min}$
at either left or right region. This is illustrated in Figure~\ref{fig:pfigb},
where the first and the second modes are turned on at the same time, 
$\chi = a_1 \chi_1^v + a_2 \chi_2^v$, and the dilaton has a cross term 
$\phi^v_{1,2}$ in \eqref{phiv123} which is an odd function of $\mu$. 
In the figure, there are three extrema which can be identified as 
the QES$_{\rm min}$, a bulge and an appetizer, respectively from the left.

Geometries with two Python's lunches can be obtained
by turning on three lowest modes \eqref{pv123}. See Figure~\ref{fig:pfigc}.
By now, it should be clear that one can generate very complicated 
geometries by suitably turning on higher modes. Figure~\ref{fig:pfigd}
is an example of this kind, where two modes
$\chi = a_{m} \chi_{m}^v + a_{n} \chi_{n}^v$ are turned on
with $m=151$ and $n=200$. In these figures, phases $\tau_n$'s are
set to be $\tau_n=\frac\pi{2n}$ so that the dilaton configuration is symmetric
under the time reversal $\tau \rightarrow -\tau$ which guarantees that
$\tau=0$ slice has extrema. One can consider, however, more general
configurations by choosing other values for $\tau_n$.

\begin{figure}[tbhp] 
\vskip-1.4cm
\centering
\includegraphics[width=0.62\textwidth]{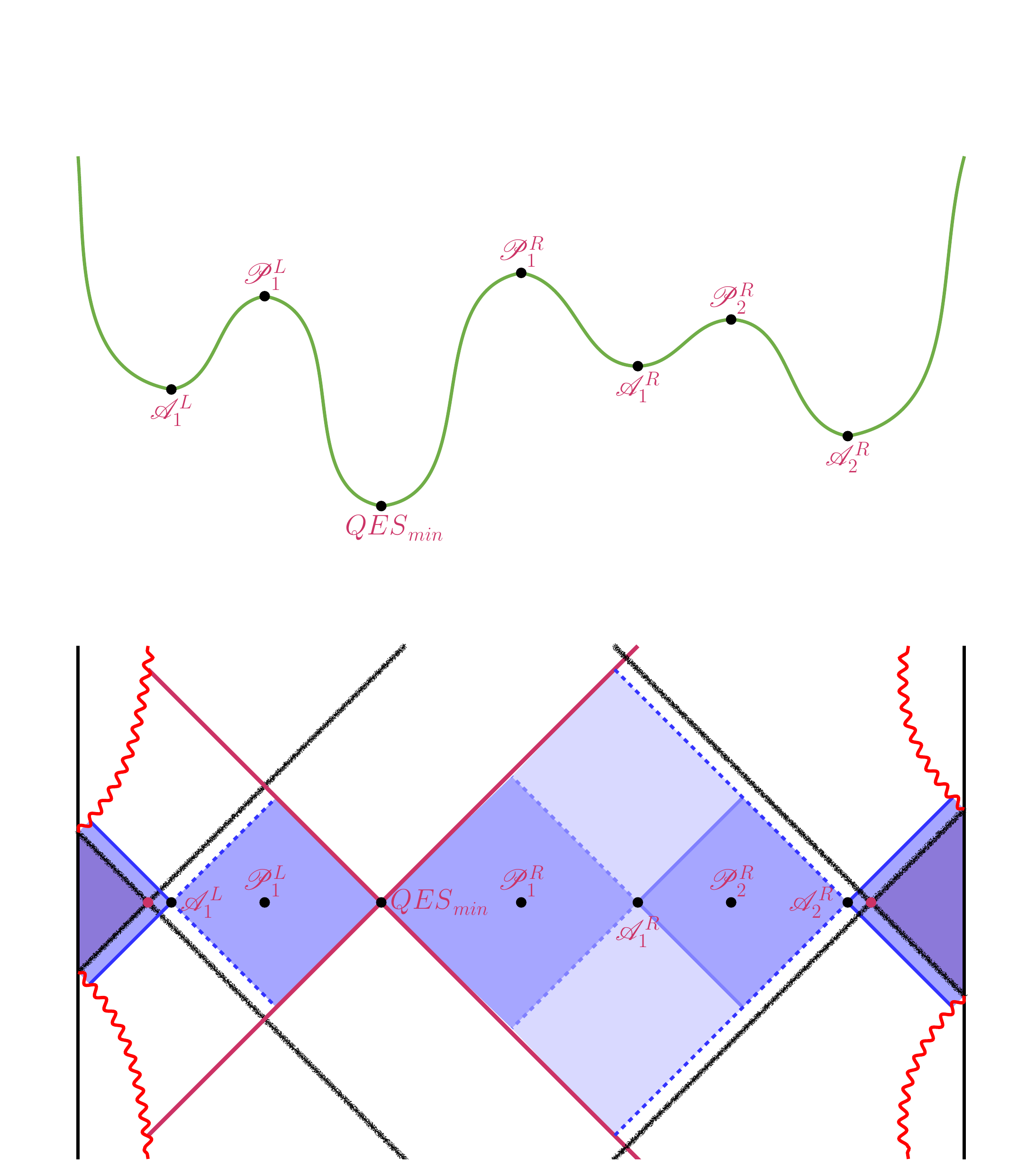}
\caption{A dilaton field with python's lunch geometries and the corresponding
Penrose diagram.}   \label{pythonfig}
\end{figure}

In Figure~\ref{pythonfig}, we draw a typical dilaton configuration
and the corresponding Penrose diagram of deformed two-sided black hole
spacetime. There is a unique QES$_{\rm min}$ at the true minimum of the dilaton. 
At the local maxima, we have three bulges denoted by 
$\mathcal{P}_1^L$, $\mathcal{P}_1^R$ and $\mathcal{P}_2^R$, respectively. 
There are also three appetizers denoted by $\mathcal{A}_1^L$, 
$\mathcal{A}_1^R$ and $\mathcal{A}_2^R$, respectively, and they correspond
to local minima. Superscript L and R refer to the relative positions
with respect to the QES$_{\rm min}$. In the Penrose diagram, horizons are drawn in black
lines. Note that all the extrema are behind the horizons as argued in 
section~\ref{sec4}. The spacetime is divided into the left and the right
wedges with QES$_{\rm min}$ as a separating point, as drawn in red lines. 
Then, according to the entanglement wedge reconstruction hypothesis, 
all local bulk operations in each wedge should be able to be reconstructed
in the corresponding boundaries as operations on each boundary Hilbert space.
In the left wedge having an appetizer $\mathcal{A}_1^L$ and a bulge
$\mathcal{P}_1^L$, the local bulk degrees of freedom in the leftmost wedge
outside $\mathcal{A}_1^L$ should admit a simple boundary
reconstruction based on the HKLL reconstruction~\cite{Hamilton:2005ju,Hamilton:2006az},
including those between the horizon and the 
appetizer~\cite{Engelhardt:2021mue}. The degrees of freedom associated 
with Python's lunch with $\mathcal{P}_1^L$ is discussed in
section~\ref{sec6}. In the right wedge, there are two Python's lunch 
geometries with bulges $\mathcal{P}_1^R$ and $\mathcal{P}_2^R$ respectively,
which are separated by the appetizer $\mathcal{A}_1^R$. The local bulk
degrees of freedom in the right wedge are then associated with
one of the Python's lunches or the rightmost wedge outside the appetizer
$\mathcal{A}_2^R$. Note that the generalized entropy of the first bulge
$\mathcal{P}_1^R$ is larger than that of the second bulge $\mathcal{P}_2^R$.

As we have demonstrated above, extrema of the dilaton can change depending
on matter perturbations. Suppose that with a suitable perturbation,
the dilaton value at, for example, $\mathcal{A}_1^R$ is lowered and 
eventually gets smaller than the dilaton value at the current QES$_{\rm min}$. 
Then the position of the QES$_{\rm min}$ would make a transition from the current
QES$_{\rm min}$ to $\mathcal{A}_1^R $. Accordingly, the degrees of freedom associated with
$\mathcal{P}_1^R$ would belong to the left wedge of the new QES$_{\rm min}$. In this case,
those degrees would no longer be reconstructed from the right boundary
according to the entanglement wedge hypothesis. Instead, it would be
reconstructed from the left boundary.

\section{Complexity, reconstruction and ghostly interactions 
}\label{sec6}

\begin{figure}[tb]
\vskip-1cm
\begin{center}
\includegraphics[width=14cm,clip]{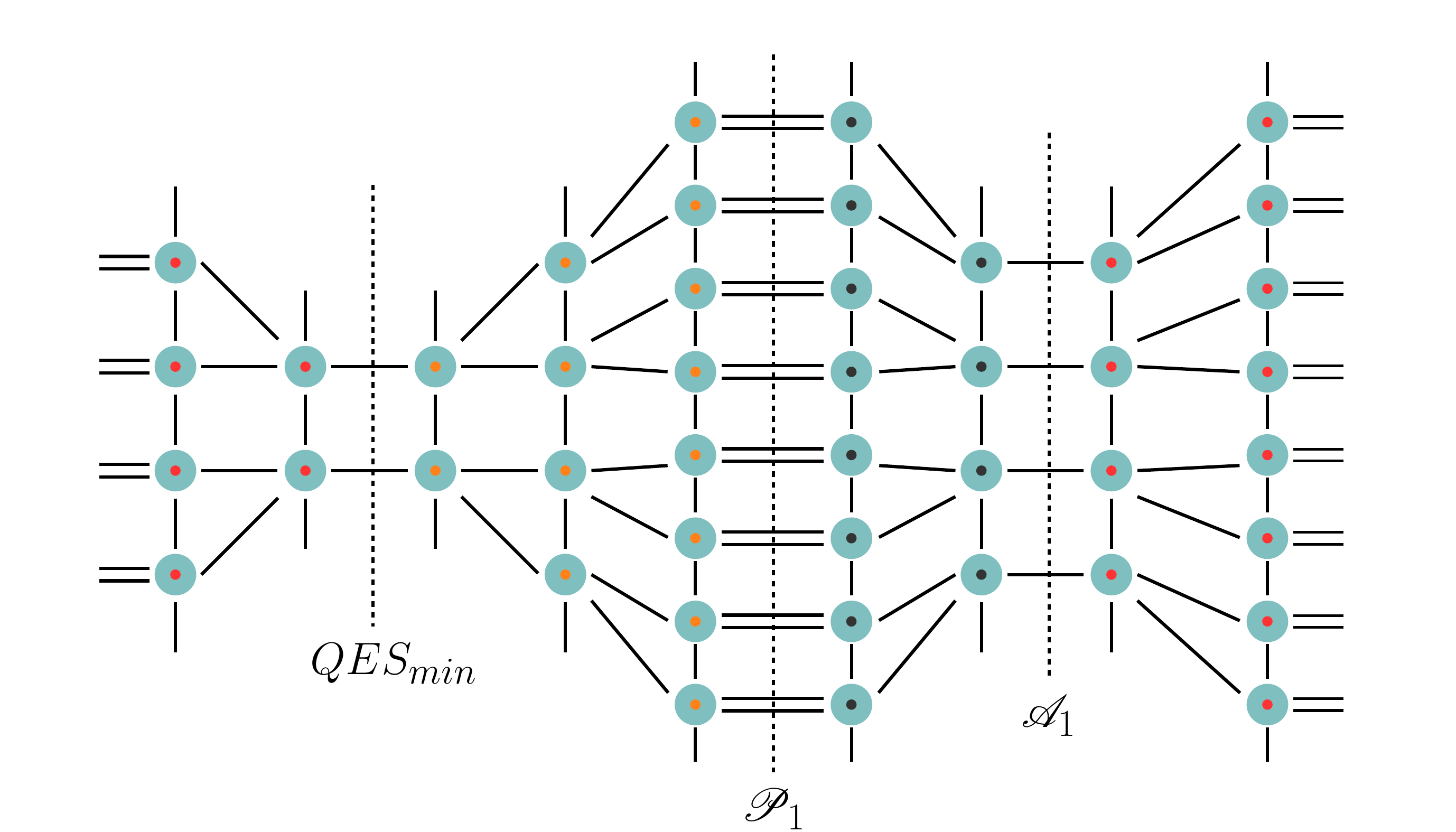}
\end{center}
\vskip-1cm
\caption{
\label{figuretensor} \small We draw the tensor network diagram of the Python's lunch geometry. In this diagram, the first and the last vertical legs in each column are identified. Each circle represent a perfect tensor with six legs. The minimal dotted cut-line represents the  QES$_{\rm min}$, the dotted line labeled by ${\cal P}_1$ the bulge  and 
 the one labeled by ${\cal A}_1$ the appetizer. The isometry $T_1$ maps from 
${\cal H}_{Q}$ to ${\cal H}_{{\cal P}_1}$. The orange colored dots between the QES$_{\rm min}$ and the bulge ${\cal P}_1$ represent  ancilla qubits required in order to make the map $T_1$ be unitary.  
The map from ${\cal H}_{{\cal P}_1}$ to ${\cal H}_{{\cal A}_1}$  can be given by an adjoint of isometry $T_2$. This requires a post-selection and the black dots represent
the post-selected qubits.  
}
\end{figure}

In this section, we would like to first briefly review  the restricted circuit complexity associated with
Python's lunch geometries \cite{Brown:2019rox,Engelhardt:2021qjs}. For definiteness, we limit our attention to the right wedge that lies on the 
right side of the true QES$_{\rm min}$ as illustrated in Figure~\ref{figuretensor} and take,  for simplicity, a basic lunch with one bulge ${\cal P}_1$ and one appetizer 
${\cal A}_1$
\footnote{At each point, the associated entropy in the classical regime is given by the formula $S(P)= \frac{\phi(P)}{4G}+ S_0$ in our JT model.}. We consider an isometry map $T$ from a state $|\psi\rangle_Q$ $( \in {\cal H}_Q$ with basis $\{ | I \rangle \} )$ defined on the  QES$_{\rm min}$ to the appetizer state $|\psi\rangle_{{\cal A}_1}$
$(\in {\cal H}_{{\cal A}_1}$ with basis $\{ | i \rangle \} )$. This requires $S_{{\cal A}_1} \ge S_{Q}$ which follows from the fact that the QES$_{\rm min}$ is the minimum of all extremal surfaces.
The bulge Hilbert space ${\cal H}_{{\cal P}_1}$ can be written as ${\cal H}_\alpha \otimes{\cal H}_{{\cal A}_1}$ where ${\cal H}_\alpha$ with basis $\{ | \alpha \rangle \} $ is associated with the Python's lunch degrees of freedom. 
The  number $n$ of its associated qubits is identified as
\be\label{alphabit}
n= \log_2 |\alpha|\ ,
\ee 
where $|\alpha|$ is the dimension of the Hilbert space ${\cal H}_\alpha $ with
$\log |\alpha|=\Delta S^{\cal P}_{11}=S_{{\cal P}_1}-S_{{\cal A}_1}$. 
The map $T$ is realized as follows. In the tensor network picture  of the  bulk  \cite{Pastawski:2015qua}
illustrated in Figure \ref{figuretensor}, one has an isometry $T_1$ from ${\cal H}_{Q}$ to ${\cal H}_{{\cal P}_1}$ but the map from ${\cal H}_{{\cal P}_1}$ to ${\cal H}_{{\cal A}_1}$ cannot be isometric. Instead it is given by the adjoint of an isometry $T_2$ and hence 
constructing the map $T=T^\dagger_2 T_1$ becomes nontrivial. Namely adding $m$ ancilla qubits with $m=\log_2 \big(|\alpha||i|/|I|\big)$, one may represent the 
map by  
\be
T|\psi\rangle_Q= \langle 0|^{\otimes n} U |0\rangle^{\otimes m}|\psi\rangle_Q\ .
\ee
 In Figure \ref{figuretensor}, the orange colored dots between the  QES$_{\rm min}$ and the bulge ${\cal P}_1$ represent  the ancilla qubits. 
 The black dots the represent required post-selected qubits, 
which are identified as the Python's lunch degrees of freedom in the above.
Since the postselection is nonunitary, the trick is to use the unitary operations of Grover search algorithm~\cite{Grover:1996rk,Grover:1997fa} whose the required number of simple operation is given by the order of $2^{\frac{n}{2}}$.
This leads to the circuit complexity of the Python's lunch 
\be
C \sim 2^{\frac{n}{2}} = e^{\frac{1}{2}\Delta S^{\cal P}_{11}}\ ,
\ee  
as was shown in \cite{Brown:2019rox}.

For a general Python's lunch, we again consider a general Python's lunch geometry in the right wedge
which involves bulges ${\cal P}_i$ and appetizers ${\cal A}_j$ with $i,j=1,2,\cdots, M$.
 In \cite{Engelhardt:2021qjs}, it was shown that the associated circuit complexity is given by
\be
C \sim  e^{\frac{1}{2}{\rm max}_{i \le j}\Delta S^{\cal P}_{ij}}\ ,
\ee
where $\Delta S^{\cal P}_{ij} =S_{{\cal P}_i}-S_{{\cal A}_j}$ defined only for $i\le j$. We shall call ${\rm max}_{i \le j}\,\Delta S^{\cal P}_{ij}$ as $\Delta S_{\cal P}$ and the relevant
bulge and appetizer of this maximum as {\it maximum} bulge and  appetizer respectively. 

Next we review the Petz map reconstruction of code space operation from the boundary theory
and its operational complexity in the presence of the Python's lunch degrees of freedom \cite{Zhao:2020wgp}.
For this purpose, we consider a code space state of   ${\cal H}_a$ (with a basis $\{ | a \rangle \} )$  that is embedded into the full Hilbert space  of ${\cal H}_\alpha \otimes {\cal H}_i $ on the right  multiplied with  the  Hilbert space ${\cal H}_{I}$ on the left where the left and the right side are entangled through the  QES$_{\rm min}$. 
Specifically this embedding defines an isometry $V$ given by
\be\label{isometryv}
|a\rangle \longrightarrow  V |a\rangle = |\psi \rangle_{RL}\ ,
\ee
where
\be\label{psirl}
 |\psi \rangle_{RL} =\frac{1}{\sqrt{|I|}} \sum_I  U |aI I_0\rangle_R|I \rangle_L\ .
\ee
This requires some clarifications; First the unitary operator   $U$ is  introduced in such a way that
\be\label{}
U |aI I_0\rangle_R =\sum_{\alpha i} |\alpha i\rangle \, U^{\alpha i}_{aI I_0}\ ,
\ee
where $|I_0\rangle$ represents a specific    ancilla qubit state and their Hilbert space basis
will be denoted by $\{ | I'_0 \rangle \} $. The number of total ancilla qubits is determined such that
\be\label{}
|a| | I| |I'_0| = |\alpha| |i|\ .
\ee
Again the $\alpha$-qubits represent the Python's lunch degrees of freedom, whose 
  number of qubits  is given by (\ref{alphabit}) as before. On the other hand, the $i$-qubits are for the degrees of freedom with relatively low-complexity, which lie on and outside the outermost appetizer extremal surface. We take their Hilbert space dimension to be large enough such that
 \be\label{}
|i| > |a| |\alpha| |I|\ .
\ee
The $I$-qubits represent the  QES$_{\rm min}$ degrees of freedom, which are responsible for the left-right entanglement in the state $|\psi\rangle_{RL}$. We also introduce a further separation of the ancilla qubits into two parts as indicated by $I_0=I_{10}I_{20}$ and their basis by  $\{ | I'_{10}I'_{20} \rangle \}$ with the relation $|i| =  |\alpha| |a||I||I'_{10}|$. With help of this setup,
we define $U$ fully by
\be\label{}
U |aI I'_0\rangle_R =\sum_{\alpha i} |\alpha i\rangle \, U^{\alpha i}_{aI I'_0}\ ,
\ee
without the restriction of the specific ancilla qubit state. Now, the operator $U^\dagger$ is completely defined as a unitary operator  acting on ${\cal H}_{\alpha}\otimes {\cal H}_{i}$.  The Python's lunch state 
$|\psi\rangle_\alpha$ is not available from the viewpoint of a right-side observer who can perform only relatively low-complexity operations. Thus from the viewpoint of this observer, one is tracing over  
the ${\cal H}_\alpha\otimes {\cal H}_{I_L}$ space in the construction of the Petz map below.

Finally  when the unitary matrix $U$ is assumed to be random enough \cite{Zhao:2020wgp},  
the operator $\tilde{U}$ defined  by
\be\label{}
\tilde{U} |\alpha aI I'_{10} I_{20}\rangle\, ( \equiv |e_{\alpha aI I'_{10}}\rangle)=\sum_{ i} | i\rangle \, U^{\alpha i}_{aI I'_{10}I_{20}}\sqrt{|\alpha|} 
\ee
becomes 
a unitary operator that is acting on ${\cal H}_{\alpha a I I'_{10}} \cong {\cal H}_{i}$ ignoring any negligibly small corrections  in all practical senses.  
Namely the set $\{ |e_{\alpha aI I'_{10}}\rangle \}$ forms an orthonormal basis of  ${\cal H}_{\alpha a I I'_{10}} \cong {\cal H}_{i}$ again ignoring any negligibly small corrections.  Note also that $\tilde{U}$ depends on a    specific $I_{20}$, which indicates the state dependence in the above construction of $\tilde{U}$.

\begin{figure}
\vskip-1cm
\begin{center}
\includegraphics[width=9cm,clip]{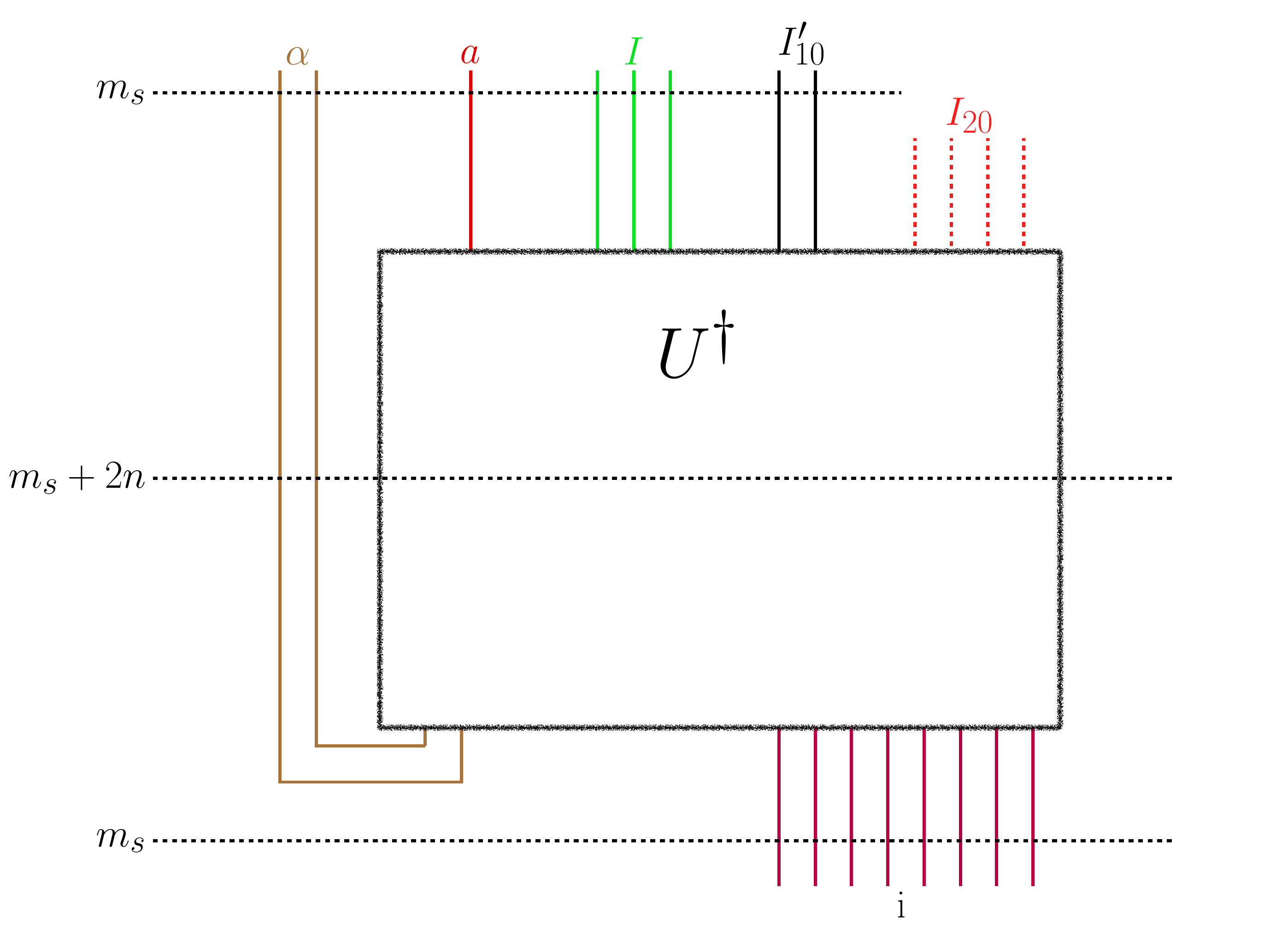}
\end{center}
\vskip-1cm
\caption{
\label{figpetz} \small We draw the ${\tilde{U}}^\dagger$ operation in terms of $U^\dagger$. 
One requires postselection of the $2n$ $I'_{20}$ qubits to the $|I_{20}\rangle$ state. These legs are denoted by dotted lines.  In addition, the number  of  the $i$-qubits is denoted by   $m_s\, (=\log_2 |i|)$.
}
\end{figure}

The 
Petz map of our interest can reconstruct a code space operation $W$, which may be understood as a black hole interior operator in the Hayden-Preskill protocol~\cite{Hayden:2007cs}, 
\be\label{}
|a\rangle \longrightarrow  W |a\rangle = \sum_b |b \rangle W_{ba}\ ,
\ee
from the state $|\psi \rangle_{RL}$ by some operation within the Hilbert space ${\cal H}_i$ alone. The map is realized by~\cite{Cotler:2017erl,Chen:2019gbt}
\be\label{}
M_P =\sigma^{-\frac{1}{2}}\, {\rm tr}_{\alpha, I_L} V W V^\dagger             \, \sigma^{-\frac{1}{2}}\ ,
\ee
with
\be\label{}
\sigma= {\rm tr}_{\alpha, I_L} V V^\dagger    \ .
\ee
One then finds that, for a specific $I_{10}$ inherited from $V$, 
\be\label{}
\sigma=\frac{1}{|\alpha||I|} \sum_{\alpha, a, I} |e_{\alpha a I I_{10}}\rangle \langle e_{\alpha a I I_{10}} |\ .
\ee
It is clear that the reconstruction is made by the operation within the space of ${\cal H}_i$.
This map may be rewritten as
\be\label{}
M_P =   \tilde{U} W    {\tilde{U}}^\dagger\,.  
\ee
As was already shown, ${\tilde U}^\dagger$ may be expressed in terms of matrix $U^\dagger$ with some postselection, which is illustrated in Figure \ref{figpetz}. 
It is also clear that the ${\tilde U}^\dagger$ action  involves a Python's lunch configuration 
of $2n$ extra qubits. In addition, its  realization by $U^\dagger$ requires
the postselection to the $|I_{20} \rangle $  state as illustrated by dotted legs in the Figure \ref{figpetz}. 
Hence we conclude that the operator complexity of the Petz map becomes
\be\label{}
C_{\rm rec} \sim 2^n = e^{\Delta S_P}\ ,
\ee
which was first shown in \cite{Zhao:2020wgp}. 

Finally we would like to discuss 
observational 
difficulties in detecting the Python's lunch degrees of freedom (see~\cite{Bouland:2019pvu} for some related discussions). 
In our 
setup,
a computationally bounded  observer performs relatively simple operations that are limited to the Hilbert space ${\cal H}_i$   introduced above. 
Of course, such probe of the Python's lunch degrees of freedom will not be feasible. 
 For instance, the bulk causality, 
in the classical regime, forbids the boundary observer to detect these degrees of freedom  since they lie behind the horizon. Hence any of these probes 
ought to be based on certain quantum effects such as Hawking radiation. 
Below we shall show that their observational signature is suppressed exponentially in $-1/G$ and, hence, nonperturbative.

As before, we use the isometry map $V$ in (\ref{isometryv}) where we take an empty code space, {\it i.e.} $\{ a\} = \emptyset$. Then the state in (\ref{psirl}) may be rewritten as
\be\label{}
 |\psi \rangle_{RL} =\frac{1}{\sqrt{|\alpha||I|}} \sum_{\alpha,I} |\alpha\rangle  |P(\alpha,I) \rangle |I \rangle_L\ ,
\ee
where the $(\alpha,I)$  state reads\footnote{
Strictly speaking, the case of  $\{ a\} = \emptyset$ was not included in the consideration of \cite{Zhao:2020wgp}.
}
\be\label{}
 |P(\alpha,I) \rangle = \sum_{i}|i \rangle {\tilde{U}}^i_{\alpha I I_{10}}= \sum_{i}|i \rangle {{U}}^{\alpha i}_{ I I_{10}I_{20}} \sqrt{|\alpha|}\ ,
\ee
with the orthonormality 
\be\label{}
\langle P(\alpha,I)  |P(\alpha',I') \rangle =\delta_{\alpha\alpha'}\delta_{I I'}\ .
\ee
One may view the $(\alpha,I)$ state as the maximally-entangled counterpart of the Python's lunch state
$|\alpha\rangle|I \rangle_L$.

We  would like to probe the state $| P(\alpha,I)  \rangle$ by a certain interaction amplitude that
induces a transition from $|i_0\rangle$ to $|j_0\rangle$ with an amplitude
$A^{j_0}_{ i_0}$ ($i_0, j_0 \in \{ i \}$). This amplitude is assumed  to be ${{\cal O} (1)}$.
We may introduce $K_0$ such 
amplitudes 
$A^{j_{(k)0}}_{i_{(k)0}}\, (k=1,2,\cdots,K_0)$ to confirm the $| P(\alpha,I)  \rangle$ 
state but let us drop the $k$ index below for the simplicity of presentation. 

The transition amplitude from  $| P(\alpha,I) \rangle$ to  $|j_0\rangle$ then becomes
 \be\label{}
A= \langle i_0 |P(\alpha,I) \rangle A^{j_0}_{i_0}\ ,
\ee
whose magnitude squared  may be
 estimated as
 \be\label{}
p=|A|^2 \sim |\langle i_0 |P(\alpha,I) \rangle|^2 \sim {\cal O}(C^{-2})\ .
\ee
To show this, we note that 
 $|P(\alpha,I) \rangle= \tilde{U}|\alpha I I_{10}\rangle$. As was just shown,
the complexity of the unitary $\tilde{U}$ (or equivalently ${\tilde{U}}^\dagger$) is given by $C_{rec}\sim e^{\Delta S_P}$. This implies
that the $\tilde{U}|\alpha I I_{10}\rangle$ state is spreading nearly uniformly
over some $C_{rec}$-number of its basis states. Therefore we conclude that
 \be\label{}
 |\langle i_0 |P(\alpha,I) \rangle| \sim {\cal O}(C_{rec}^{-1/2}) \sim {\cal O}(C^{-1})\ .
\ee
Thus the amplitude is indeed exponentially suppressed in $-1/G$ and  the $(\alpha,I)$  
state cannot be detected through any perturbative interactions on the boundary. Namely the Python's lunch degrees
lie within the almost completely dark sector of the system. This is also consistent with the claim \cite{Kim:2020cds} that any ghost operators acting on
the Python's lunch Hilbert space ${\cal H}_\alpha$ nearly commute with  any operation on ${\cal H}_i$ applied by any computationally bounded observer.  As in \cite{Kim:2020cds}, our argument for the above exponential suppression involves 
a state dependence since the $P(\alpha,I)$ construction depends on $I_{10}$ and $I_{20}$. 

Finally we would like to comment on the degrees of freedom lying outside the outermost appetizer but still inside the horizon. The restricted complexity $\tilde{C}_V$ associated with these degrees of freedom may be  estimated as the volume of the region divided by ${G\ell}$ following for instance  the volume conjecture \cite{Susskind:2014rva}
of holographic complexity.
Repeating a similar reasoning as in the above, the observational probability through some boundary ${\cal O}(1)$ interactions  may be estimated as
\be
{\tilde{p}} \sim  {\cal O} ({\tilde{C} }^{-1}_{V}) \sim {\cal O}(G)\,.
\ee
Hence these degrees of freedom can be probed perturbatively by appropriate boundary  experiments.
This is consistent with the fact that, in evaporating black holes, information on such degrees of 
freedom may be transferred to their Hawking radiation after their Page time, which is a semiclassical effect of ${\cal O}(G)$. This is also consistent with the observation \cite{Engelhardt:2021mue} that the reconstruction of such degrees of freedom may be 
achieved with relatively simple boundary operations.


\section{Conclusions  
}\label{sec7}

In this work, we have presented the deformation of the two-dimensional black hole by a massless scalar field which realizes  Python's lunch geometries. First, we have studied the deformation of JT gravity by the massless scalar field. With general source and state deformations, a full general solution of the dilaton has been derived. And then, from the asymptotic behavior of the dilaton solution, we have identified the locations of the deformed horizons and the  future/past infinities. Here, we have shown that the difference of the future infinity of one side and the past  infinity of the other side is less than $\pi$ in the global time coordinate, and from the NEC we have also demonstrated that any extremal points of the dilaton are located behind of the horizon. In addition, we have calculated the extremal points of the deformed dilaton solution to find the QES$_{\rm min}$, Python's lunches and appetizers. We have shown that the position and the number of Python's lunches and appetizers can vary drastically by tuning the deformation parameters. Finally, we have discussed the restricted complexity of the general Python's lunches and the reconstruction of a code space operator via a Petz map. We have explained why it is difficult to probe the Python's lunch degrees of freedom and have shown that  the observational probability of Python's lunch degrees of freedom  is exponentially 
suppressed.

The undeformed eternal black hole can be holographically described by the TFD state 
\begin{equation} \label{}
|{\rm TFD}\rangle\,=\, {1\over \sqrt{Z}} \sum_n e^{-{\beta E_n\over 2}} |E_n\rangle_L |E_n\rangle_R\,, \nonumber
\end{equation}
where $Z$ is a normalization factor.
%
The deformation by a massless scalar field produces an excitation of the black hole on its left and right sides asymmetrically, which results in the holographic dual state to the Python's lunch geometry of which the left and right sides are asymmetric in general. To the leading order in source and vev deformations, 
the corresponding deformations of the TFD state can be worked out; See \cite{Bak:2017xla} for the detailed construction of such 
states in a three dimensional AdS gravity with a scalar field. 
It would be interesting to analyze the entanglement structure of the deformed state and to understand how the Python degrees of freedom are incorporated in the state. We leave this general construction of the deformed  state 
dual to Python's lunches for future works.

It is well-known that the JT gravity for the nearly-AdS$_2$ can be reduced to a Schwarzian theory on the boundary. In this reduction, the scalar field in the AdS bulk will be coupled to the Schwarzian mode on the boundary. Under the deformation by the massless scalar field, the Schwarzian action will be also deformed in a way that the left and right Schwarzian modes are now asymmetric. In this deformed Schwarzian theory, many questions are left unanswered. How does the Python's lunch appear in this deformed Schwarzian theory? How can we evaluate the complexity and  estimate the difficulty in detecting the Python's lunch degrees of freedom? We also leave those questions for future studies.


\subsection*{Acknowledgement}
We would like to thank Andreas Gustavsson for careful reading of the manuscript.
D.B. was
supported in part by
NRF 
2020R1A2B5B01001473 and by  Basic Science Research Program
through 
NRF funded by the Ministry of Education
(2018R1A6A1A06024977). 
 C.K.\ was supported in part by NRF 
2019R1F1A1059220.
S.-H.Y. was supported  in part by 
NRF 
2018R1D1A1A09082212 and NRF 
2021R1A2C1003644 and supported  by Basic Science Research Program through the NRF funded by the Ministry of Education 2020R1A6A1A0304787.
JY was supported by 
NRF 
(2019R1F1A1045971, 2022R1A2C1003182)
and by an appointment to the JRG Program at the APCTP through the Science and Technology Promotion Fund and Lottery Fund of the Korean Government. This is also supported by the Korean Local Governments - Gyeongsangbuk-do Province and Pohang City.



\begin{thebibliography}{99}\label{bib}



\bibitem{Bekenstein:1973ur}
J.~D.~Bekenstein,
``Black holes and entropy,''
Phys. Rev. D \textbf{7}, 2333-2346 (1973)

\bibitem{Hawking:1974rv}
S.~W.~Hawking,
``Black hole explosions,''
Nature \textbf{248}, 30-31 (1974)

\bibitem{Bombelli:1986rw}
L.~Bombelli, R.~K.~Koul, J.~Lee and R.~D.~Sorkin,
``A Quantum Source of Entropy for Black Holes,''
Phys. Rev. D \textbf{34}, 373-383 (1986)

\bibitem{Srednicki:1993im}
M.~Srednicki,
``Entropy and area,''
Phys. Rev. Lett. \textbf{71}, 666-669 (1993)
[arXiv:hep-th/9303048 [hep-th]]


\bibitem{Page:1993df}
D.~N.~Page,
``Average entropy of a subsystem,''
Phys. Rev. Lett. \textbf{71}, 1291-1294 (1993)
[arXiv:gr-qc/9305007 [gr-qc]].

\bibitem{Hawking:1974sw}
S.~W.~Hawking,
``Particle Creation by Black Holes,''
Commun. Math. Phys. \textbf{43}, 199-220 (1975)
[erratum: Commun. Math. Phys. \textbf{46}, 206 (1976)]

\bibitem{Mathur:2009hf}
S.~D.~Mathur,
``The Information paradox: A Pedagogical introduction,''
Class. Quant. Grav. \textbf{26}, 224001 (2009)
[arXiv:0909.1038 [hep-th]].

\bibitem{Almheiri:2012rt}
A.~Almheiri, D.~Marolf, J.~Polchinski and J.~Sully,
``Black Holes: Complementarity or Firewalls?,''
JHEP \textbf{02}, 062 (2013)
[arXiv:1207.3123 [hep-th]].


\bibitem{Wall:2012uf}
A.~C.~Wall,
``Maximin Surfaces, and the Strong Subadditivity of the Covariant Holographic Entanglement Entropy,''
Class. Quant. Grav. \textbf{31}, no.22, 225007 (2014)
[arXiv:1211.3494 [hep-th]].

\bibitem{Engelhardt:2014gca}
N.~Engelhardt and A.~C.~Wall,
``Quantum Extremal Surfaces: Holographic Entanglement Entropy beyond the Classical Regime,''
JHEP \textbf{01}, 073 (2015)
[arXiv:1408.3203 [hep-th]].

\bibitem{Akers:2019lzs}
C.~Akers, N.~Engelhardt, G.~Penington and M.~Usatyuk,
``Quantum Maximin Surfaces,''
JHEP \textbf{08}, 140 (2020)
[arXiv:1912.02799 [hep-th]].


\bibitem{Ryu:2006ef}
S.~Ryu and T.~Takayanagi,
``Aspects of Holographic Entanglement Entropy,''
JHEP \textbf{08}, 045 (2006)
[arXiv:hep-th/0605073 [hep-th]].

\bibitem{Ryu:2006bv}
S.~Ryu and T.~Takayanagi,
``Holographic derivation of entanglement entropy from AdS/CFT,''
Phys. Rev. Lett. \textbf{96}, 181602 (2006)
[arXiv:hep-th/0603001 [hep-th]].

\bibitem{Hubeny:2007xt}
V.~E.~Hubeny, M.~Rangamani and T.~Takayanagi,
``A Covariant holographic entanglement entropy proposal,''
JHEP \textbf{07}, 062 (2007)
[arXiv:0705.0016 [hep-th]].



\bibitem{Almheiri:2020cfm}
A.~Almheiri, T.~Hartman, J.~Maldacena, E.~Shaghoulian and A.~Tajdini,
``The entropy of Hawking radiation,''
Rev. Mod. Phys. \textbf{93}, no.3, 035002 (2021)
[arXiv:2006.06872 [hep-th]].

\bibitem{Brown:2019rox}
A.~R.~Brown, H.~Gharibyan, G.~Penington and L.~Susskind,
``The Python\textquoteright{}s Lunch: geometric obstructions to decoding Hawking radiation,''
JHEP \textbf{08}, 121 (2020)
[arXiv:1912.00228 [hep-th]].

\bibitem{Swingle:2009bg}
B.~Swingle,
``Entanglement Renormalization and Holography,''
Phys. Rev. D \textbf{86}, 065007 (2012)
[arXiv:0905.1317 [cond-mat.str-el]].


\bibitem{Czech:2012bh}
B.~Czech, J.~L.~Karczmarek, F.~Nogueira and M.~Van Raamsdonk,
``The Gravity Dual of a Density Matrix,''
Class. Quant. Grav. \textbf{29}, 155009 (2012)
[arXiv:1204.1330 [hep-th]].

\bibitem{Headrick:2014cta}
M.~Headrick, V.~E.~Hubeny, A.~Lawrence and M.~Rangamani,
``Causality \& holographic entanglement entropy,''
JHEP \textbf{12}, 162 (2014)
[arXiv:1408.6300 [hep-th]].

\bibitem{Almheiri:2014lwa}
A.~Almheiri, X.~Dong and D.~Harlow,
``Bulk Locality and Quantum Error Correction in AdS/CFT,''
JHEP \textbf{04}, 163 (2015)
[arXiv:1411.7041 [hep-th]].



\bibitem{Jafferis:2015del}
D.~L.~Jafferis, A.~Lewkowycz, J.~Maldacena and S.~J.~Suh,
``Relative entropy equals bulk relative entropy,''
JHEP \textbf{06}, 004 (2016)
[arXiv:1512.06431 [hep-th]].

\bibitem{Dong:2016eik}
X.~Dong, D.~Harlow and A.~C.~Wall,
``Reconstruction of Bulk Operators within the Entanglement Wedge in Gauge-Gravity Duality,''
Phys. Rev. Lett. \textbf{117}, no.2, 021601 (2016)
[arXiv:1601.05416 [hep-th]].

\bibitem{Faulkner:2017vdd}
T.~Faulkner and A.~Lewkowycz,
``Bulk locality from modular flow,''
JHEP \textbf{07}, 151 (2017)
[arXiv:1704.05464 [hep-th]].

\bibitem{Cotler:2017erl}
J.~Cotler, P.~Hayden, G.~Penington, G.~Salton, B.~Swingle and M.~Walter,
``Entanglement Wedge Reconstruction via Universal Recovery Channels,''
Phys. Rev. X \textbf{9}, no.3, 031011 (2019)
[arXiv:1704.05839 [hep-th]].

\bibitem{Chen:2019gbt}
C.~F.~Chen, G.~Penington and G.~Salton,
``Entanglement Wedge Reconstruction using the Petz Map,''
JHEP \textbf{01}, 168 (2020)
[arXiv:1902.02844 [hep-th]].

\bibitem{Susskind:2014rva}
L.~Susskind,
``Computational Complexity and Black Hole Horizons,''
Fortsch. Phys. \textbf{64}, 24-43 (2016)
[arXiv:1403.5695 [hep-th]].

\bibitem{Brown:2015lvg}
A.~R.~Brown, D.~A.~Roberts, L.~Susskind, B.~Swingle and Y.~Zhao,
``Complexity, action, and black holes,''
Phys. Rev. D \textbf{93}, no.8, 086006 (2016)
[arXiv:1512.04993 [hep-th]].

\bibitem{Brown:2015bva}
A.~R.~Brown, D.~A.~Roberts, L.~Susskind, B.~Swingle and Y.~Zhao,
``Holographic Complexity Equals Bulk Action?,''
Phys. Rev. Lett. \textbf{116}, no.19, 191301 (2016)
[arXiv:1509.07876 [hep-th]].

\bibitem{Hayden:2007cs}
P.~Hayden and J.~Preskill,
``Black holes as mirrors: Quantum information in random subsystems,''
JHEP \textbf{09}, 120 (2007)
[arXiv:0708.4025 [hep-th]].

\bibitem{Harlow:2013tf}
D.~Harlow and P.~Hayden,
JHEP \textbf{06}, 085 (2013)
[arXiv:1301.4504 [hep-th]].

\bibitem{Jackiw:1984je}
R.~Jackiw,
``Lower Dimensional Gravity,''
Nucl. Phys. B \textbf{252}, 343-356 (1985)

\bibitem{Teitelboim:1983ux}
C.~Teitelboim,
``Gravitation and Hamiltonian Structure in Two Space-Time Dimensions,''
Phys. Lett. B \textbf{126}, 41-45 (1983)

\bibitem{Almheiri:2014cka}
A.~Almheiri and J.~Polchinski,
``Models of AdS$_{2}$ backreaction and holography,''
JHEP \textbf{11}, 014 (2015)
[arXiv:1402.6334 [hep-th]].

\bibitem{Nayak:2018qej}
P.~Nayak, A.~Shukla, R.~M.~Soni, S.~P.~Trivedi and V.~Vishal,
``On the Dynamics of Near-Extremal Black Holes,''
JHEP \textbf{09}, 048 (2018)
[arXiv:1802.09547 [hep-th]].


\bibitem{Maldacena:2016upp}
J.~Maldacena, D.~Stanford and Z.~Yang,
``Conformal symmetry and its breaking in two dimensional Nearly Anti-de-Sitter space,''
PTEP \textbf{2016}, no.12, 12C104 (2016)
[arXiv:1606.01857 [hep-th]].


\bibitem{Bak:2007jm}
D.~Bak, M.~Gutperle and S.~Hirano,
JHEP \textbf{02} (2007), 068
[arXiv:hep-th/0701108 [hep-th]];
D.~Bak, M.~Gutperle and A.~Karch,
JHEP \textbf{12}, 034 (2007)
[arXiv:0708.3691 [hep-th]].




\bibitem{Bak:2018txn}
D.~Bak, C.~Kim and S.~H.~Yi,
``Bulk view of teleportation and traversable wormholes,''
JHEP \textbf{08}, 140 (2018)
[arXiv:1805.12349 [hep-th]].

\bibitem{Breitenlohner:1982jf}
P.~Breitenlohner and D.~Z.~Freedman,
``Stability in Gauged Extended Supergravity,''
Annals Phys. \textbf{144}, 249 (1982)

\bibitem{Maldacena:2018lmt}
J.~Maldacena and X.~L.~Qi,
``Eternal traversable wormhole,''
[arXiv:1804.00491 [hep-th]].





\bibitem{Hawking:1973uf}
S.~W.~Hawking and G.~F.~R.~Ellis,
``The Large Scale Structure of Space-Time,''


\bibitem{Bousso:2015mna}
R.~Bousso, Z.~Fisher, S.~Leichenauer and A.~C.~Wall,
``Quantum focusing conjecture,''
Phys. Rev. D \textbf{93}, no.6, 064044 (2016)
[arXiv:1506.02669 [hep-th]].

\bibitem{Engelhardt:2021qjs}
N.~Engelhardt, G.~Penington and A.~Shahbazi-Moghaddam,
``Finding Pythons in Unexpected Places,''
[arXiv:2105.09316 [hep-th]].




\bibitem{Bak:2017xla}
D.~Bak, C.~Kim, K.~K.~Kim and J.~P.~Song,
``Holographic Micro Thermofield Geometries of BTZ Black Holes,''
JHEP \textbf{06}, 079 (2017)
[arXiv:1704.01030 [hep-th]].

\bibitem{Hamilton:2005ju}
A.~Hamilton, D.~N.~Kabat, G.~Lifschytz and D.~A.~Lowe,
Phys. Rev. D \textbf{73}, 086003 (2006)
[arXiv:hep-th/0506118 [hep-th]].

\bibitem{Hamilton:2006az}
A.~Hamilton, D.~N.~Kabat, G.~Lifschytz and D.~A.~Lowe,
Phys. Rev. D \textbf{74}, 066009 (2006)
[arXiv:hep-th/0606141 [hep-th]].

\bibitem{Engelhardt:2021mue}
N.~Engelhardt, G.~Penington and A.~Shahbazi-Moghaddam,
``A World without Pythons would be so Simple,''
[arXiv:2102.07774 [hep-th]].



\bibitem{Pastawski:2015qua}
F.~Pastawski, B.~Yoshida, D.~Harlow and J.~Preskill,
``Holographic quantum error-correcting codes: Toy models for the bulk/boundary correspondence,''
JHEP \textbf{06}, 149 (2015)
[arXiv:1503.06237 [hep-th]].



\bibitem{Grover:1996rk}
L.~K.~Grover, ``A Fast quantum mechanical algorithm for database search,'' [arXiv:quant-ph/9605043 [quant-ph]].


\bibitem{Grover:1997fa}
L.~K.~Grover, ``Quantum mechanics helps in searching for a needle in a haystack,''
Phys. Rev. Lett. \textbf{79}, 325-328 (1997)
[arXiv:quant-ph/9706033 [quant-ph]].



\bibitem{Zhao:2020wgp}
Y.~Zhao,
``Petz map and Python\textquoteright{}s lunch,''
JHEP \textbf{11}, 038 (2020)
[arXiv:2003.03406 [hep-th]].

\bibitem{Bouland:2019pvu}
A.~Bouland, B.~Fefferman and U.~Vazirani,
``Computational pseudorandomness, the wormhole growth paradox, and constraints on the AdS/CFT duality,''
[arXiv:1910.14646 [quant-ph]].

\bibitem{Kim:2020cds}
I.~Kim, E.~Tang and J.~Preskill,
``The ghost in the radiation: Robust encodings of the black hole interior,''
JHEP \textbf{06}, 031 (2020)
[arXiv:2003.05451 [hep-th]].








\end{thebibliography}
\end{document}